# *Homo denisova*, Correspondence Spectral Analysis, Finite Sites Reticulate Hierarchical Coalescent Models and the Ron Jeremy Hypothesis

Peter J. Waddell[12], Jorge Ramos[3], and Xi Tan[2]

pwaddell@purdue.edu

[1]Department of Biological Sciences, Purdue University, West Lafayette, IN 47906, U.S.A.
[2]Department of Computer Science, Purdue University, West Lafayette, IN 47906, U.S.A
[2]Department of Physics, Purdue University, West Lafayette, IN 47906, U.S.A

.

This article shows how to fit reticulate finite and infinite sites sequence spectra to aligned data from five modern human genomes (San, Yoruba, French, Han and Papuan) plus two archaic humans (Denisovan and Neanderthal), to better infer demographic parameters. These include interbreeding between distinct lineages. Major improvements in the fit of the sequence spectrum are made with successively more complicated models. Findings include some evidence of a male biased gene flow from the Denisova lineage to Papuan ancestors and possibly even more archaic gene flow. It is unclear if there is evidence for more than one Neanderthal interbreeding, as the evidence suggesting this largely disappears when a finite sites model is fitted.

**Keywords**: *Homo neanderthalensis, Homo denisova,* phylogenetic spectral analysis, Hadamard conjugation, correspondence analysis, hierarchical coalescence models, interbreeding, hybridization population genetics

# 1 Introduction

The year 2010 marked the publication in Nature and Science Magazine of the first large amounts of genomic data from archaic humans. In May 2010, Green et al. (2010) published a partial Neanderthal Genome, along with analyses that suggested there was 2-5% interbreeding between Neanderthals and the ancestors of the modern people that left Africa. Shortly after, it was announced that the mitochondrial genomic sequence of a bone from the Denisova cave in Siberia showed another lineage of archaic humans had survived in Eurasia until the time that modern humans started to settle the area (Krause et al. 2010). Finally, Christmas 2010 was marked by the publication of a set of data and analyses that strongly suggested that modern humans had interbreed with at least two distinct lineages of archaic hominids (Reich et al. 2010). Remarkably, their data set now had a geographical diverse array of modern human and two geographically widespread archaic human genomes (a Neanderthal and a Denisovan).

The analyses suggested that while the Denisovan sequence was closest to that of Neanderthals, it was a highly distinct lineage. Further, it appeared that the modern Papuan sequence analyzed had received approximately 5% of its genome from this archaic hominid, in addition to all the out of Africa genomes sharing 2-5% Neanderthal genes (Reich et al. 2010, Reich et al. 2011). Remarkably, the fraction of Neanderthal genes in all the out of African sequences was very similar. There was no evidence that the relatively gracile Han Chinese person sequenced had any fewer Neanderthal genes that the French person sequenced, indeed if anything, they seemed to show more such genes.

Apart from giving us a unique and exciting insight to our origins, the data threw the spotlight once again on the gaps in our existing methodology for analyzing genomic sequences. The genomes mostly seemed to follow a single species tree, but there was not enough time for nearly complete inbreeding or coalescence along edges of this species tree. Therefore gene trees, or more accurately, unlinked regions of the genome, could follow branching patterns different to that of the species tree. Further, there seemed to be distinct cases of interbreeding between lineages of the main species tree, which is a tricky issue for phylogenetic analysis. Finally, the data suggested rather high rates of sequencing error in the various genome sequences, up to approximately half of all the polymorphisms in some of the genomes gathered appeared to be sequencing errors (Reich et al. 2010, particularly supplement parts 2 and 3). In addition, there would also be mistaken cases of the true identity of the ancestral state as the Chimpanzee outgroup was about 14 million years separating the human sequences from the Chimpanzee outgroup. Unfortunately, no full likelihood based analysis combines all these features, although methods developed in Waddell (1995) come close, showing how to combine finite sites models with hierarchal coalescent models, and showing also how to deal with reticulate evolution.

It is therefore important to look in detail at the analysis of this data and how it might be improved. For example, some of the statistics used in Reich et al. (2010) are a redevelopment of earlier published methods, such as those in Waddell et al. (1995, 2000 and 2001). The E-statistics of Reich et al. for quartets are essentially the same as the binomial P statistics for testing disagreement of three taxon rooted trees under a coalescent model (e.g., Waddell et al. 2001). These statistics are based on four taxon patterns AABB, ABAB and ABBA, which in the rooted triplet form are 011, 101 and 110 since the outgroup with state 0 is always the left most taxon (thus giving the patterns 0011, 0101 and 0110, Waddell et al. 2001). Indeed, many of the statistics explored by Reich et al. (2010) are subsets of the more general spectral methods developed in phylogenetics by authors such as Hendy and Penny (1990), with a particular connection to population genetics as described in Waddell (1995). Reduced sub-spectral methods such as those in Waddell et al. (2000, 2001) and Reich et al. (2010) are powerful methods for probing complex population histories, including hybridization, but they ultimately lack power compared to using the full spectrum of edge lengths/substitution patterns.

Looking at the spectra of multiple groups includes the need to recognize the fact that some sites in the sequence may follow a different history to others. This is often called the "gene-tree species tree problem." However, this focus on genes may be unfortunate and can lead to some fundamental misunderstandings along with misleading estimates. A number of authors independently developed methods of estimating the ancestral population size of different species using DNA sequences. The method's of Waddell (1995) which led to the first robust estimates of the ancestral population size of humans and chimps were inspired by direct communication with Dick Hudson, one of the originators of coalescent theory. This included discussion the seminal paper, Hudson (1992), that pointed out the need to consider not only the probability that a stretch of DNA follows a particular history, but also the probability of observing it (which is a function of the mean edge lengths of the different gene trees as well as the probability of those trees). In addition, the probability of observing a particular "gene" tree, is dependent upon how many potentially identically segregating sites it includes, which is, in turn a function of the local recombination rate. Such concepts fit in naturally with the conceptual framework of the Hadamard conjugation (Hendy and Penny 1993). In these methods, single site patterns are the object of interest. This in turn leads to a certain freedom from the tyranny of being unable to know where precisely recombination has occurred, since a single orthologous site may be correlated with it's neighbor, but is itself indivisible (given a reliable alignment).

Independent attempts to study ancestral polymorphism (e.g., Ruvolo 1997) tend to infer population history by counting how many trees of different types are reconstructed by different genes. This is unfortunately a misleading method given unequal numbers of potentially segregating sites and/or recombination (Waddell 1995). Directly counting "gene trees" is an inherently unreliable and biased-method. Indeed, such methods may be called "Dick-less" since they do not take to heart the essential insights of Hudson (1992) on link between what is observed and what the underlying genetic model is. In contrast, spectral methods allow an unbiased reconstruction of the polymorphism present in ancestral species (Waddell 1995, see also Waddell et al. 2000, 2001, 2002). Such approaches are very much the "Full-Monty" (an English term meaning "the full deal") as they can be extended to include a variety of ways to correct for multiple changes at a site due to ongoing mutation. In this way they are able to give unbiased estimates of ancient ancestral diversity and hence to escape the bias created by spectral methods based purely on the infinite-sites model (Waddell 1995). With much going for this approach, it does have one weakness, that is, it can be difficult to estimate the precise variance of the estimates, since it is not determinable precisely where recombination has occurred (and hence how many genetic "units" or loci are involved). This problem is addressed in this article.

Finally, we consider the old question of what a species is, particularly as it pertains to humans. It is generally accepted by modern biologists that living humans comprise a single species (e.g., Waddell and Penny 1996), yet it is also increasingly recognized that there is no single essence to a species. The opinion that all living humans are a single species is based on a variety of criteria including evidence for low genetic differentiation and a surprisingly recent time of origin (e.g., Penny et al. 1995,). Further, it is also often argued that extinct populations, such as Neanderthals, might best be considered a subspecies or even a population of modern humans. The latter view is reinforced by recent findings that Neanderthal genes have apparently entered the whole no African modern gene pool, at levels of 2-5% (Green et al. 2010).

On the other hand, human groups show a strong cultural capacity to self-identify. This may include active genocide against groups not considered "like." A well-studied example of this was the extermination of Tasmanian aborigines by European Australians. Such a powerful ability to nearly completely replace the gene pool of initially numerically superior competing populations would seem to generally be more a property of different species than members of the same species. So far it is unclear whether this property is readily found within other taxa called species, including the other apes. This is an area that genomics is now capable of unraveling.

Acting as devil's advocate is an important part of scientific process and in that spirit, here we may treat each human population sampled as a different "species." As one esteemed Anglo-Saxon American geneticist colleague commented on hearing some of these arguments, he was not sure about the other groups, but could see how the French could be a different species, i.e., *Homo froggieii*.

*Thus we find ourselves in the close company of Darwin when he asked "The question whether mankind consists of one or several species (Descent of Man, 1871, p. 24)" saying "it is a hopeless endeavor to decide this point on sound grounds, until some definition of the term "species" is generally accepted; and the definition must not include an element that cannot possibly be ascertained, such as an act of creation."*

A concept of species that seems particularly apt to test in describing hominid evolution is the ability of a smaller group to replace a large population of close relatives through direct competition with minimal genetic exchange. This an ultimate form of the expression of a selfish genome. Darwin predicted this for the human future, writing *"At some future point, not distant as measured by centuries, the civilized races of man will almost certainly exterminate and replace the savage races throughout the world"* (here, civilized referred to European cultures). While this now hopefully seems less likely to occur, the era of genomic data is allowing us to address questions on matters of the closure of gene pools, such as which hominid groups have acted more like different species, when compared to other taxa of our own order Primates.

## 2  Materials and Methods

The data come directly from Reich et al. (2010). The processing of this data is rewritten here as best we understand it based on discussions/correspondence with the authors (particularly Nick Patterson and Martin Kircher) and details largely in the supplement to their article (for example, p32). The chimp reference genome is used as the outgroup to call the ancestral state. This genome comprised the super-contigs of 23 autosomes (the same as human except human chromosome 2 is represented as two separate chromosomes 2a and 2b, as in the chimp). In addition, the chimp reference genome contains collections of contigs that are located to particular chromosomes, but do not fit well into the known super-contigs of those chromosomes. These are called random blocks and also serve as targets for alignments of the hominid genomic reads. Next, independently, the short reads of five modern humans and the Denisova specimen were aligned to this genomic target using the BWA aligner. Only reads showing a high confidence of a close and unique mapping were kept. The reads of the three Vindija Neanderthals were pooled and these were mapped in the same basic way except that the aligner used was ANFO as it was deemed to have a better capacity to deal with the unique sequencing error spectrum of the Neanderthal data (Reich et al. 2010).

A script was then written (by N. Patterson) to read along these alignments. Whenever a site in chimp was aligned to at least one read of each of the other seven taxa (the five modern humans, *Homo san* or S, *Homo yoruba*, Y, *Homo froggieii*, F, *Homo han*, H, and *Homo papuan*, P, *H. denisova,* D*,* and *H. neanderthalensis,* N) that met their specified quality standards was the site pattern kept. If a taxon had more than one read, the read to represent that taxon was randomly chosen. Only transversional changes were kept which is equivalent to recoding the data as R = A or G and Y = C or T. This is a standard approach in spectral analysis (e.g., Hendy and Penny 1993) and ensures robustness as (1) transversions occur less frequency than transitions in these sequences, reducing the rate of double mutations at an aligned site (2) Many of the sequencing errors in the Neanderthal in particular are transitional changes. Despite this, perhaps up to half of all the base changes in this data are sequencing errors (Reich et al. 2010, SI2 and SI 6 in particular).

The results were written out for each chromosome separately and also separately for chunks of the random blocks from each chromosome. These are the data that underlie the analyses of Reich et al. SI6-8 and figure 3 in the main article (plus the X chromosome which was not reported in that section, but was investigated elsewhere, e.g. SI2). The indexing used was basically identical to the binary numbering of Hendy and Penny (1993), except they were written left to right rather than right to left. For example, the original ordering used was FHPSYNDC. We reorder these to CDNSYFHP so that they confirm to the standard binary ordering where the outgroup (state 0) is always on the left and so that they follow a linearization of the expected "species" tree, which is (C,((D,N),(S,(Y,(F,(H,P)))))) (such reordering has immense advantages in remembering and interpreting this type of data). Thus, a pattern where D and P alone share the derived state (pattern DP) is expressed as the binary number 01000001. This binary number is $2^{i-1} + 2^{j-1}$ (where $i$ is the index of P (1 in this case) and $j$ is the index of D (7)), which is $2^{1-1} + 2^{7-1} = 1 + 64$, or 65, which is it's base 10 index.

The analyses of these raw counts, kindly communicated by Nick Patterson, began with exploration with $X^2$ homogeneity statistics (using Microsoft Excel and R, Ihaka and Gentleman 1996, R Core Development Team 2005). These analyses were then duplicated (and automated) with PERL scripts. In particular, the deviations of patterns frequencies from mean values were inferred by looking at the pairwise $X^2$ distance between all chromosomes (and their grouped random chromosome blocks), between all patterns across chromosomes (with or without random blocks included) and between each pattern on each chromosome with it's expected frequency based on independence (the product of the row total times the column total, divided by the total number of patterns, a type of homogeneity test). These were further explored and visualized by building trees using PAUP* (Swofford 2000, versions b108 to a122) and building trees and networks using SplitsTree4 (Huson and Bryant 2006). Correspondence analysis was performed in R using the package CA (Ihaka and Gentleman 1996, R Core Development Team 2005), along with a variety of modifications and extensions we made to the visualization of the plots.

Analyses based on a coalescent model for eight species (represented as a seven taxon rooted tree with Chimp serving as the outgroup) were based on an Excel spreadsheet. The calculations in this spreadsheet are a generalization of those in Hudson (1993). That is, the excepted internal edge lengths of each type of gene tree were calculated under a Wright-Fisher infinite sites model using integration (Maple was used to check the form of the integrals). Then, after multiplication of each expected internal edge length by the relative probability of that type of gene tree, the sum of the expected length all internal edges of a particular index was made. This yields the expected frequency of that bipartition (or substitution pattern) based on the assumptions of an infinite sites model. These in turn are the basis for estimates of hierarchically structures coalescent models, with or without correction for multiple substitutions per site (Hudson 1992, Waddell 1995). The correction for multiple substitutions per site is generally based on a first order Markov transition matrix. In theory, this should be made over every single realization of a gene tree (and for every single site rate if some sites are assumed to mutate more than others), but it can be simplified with a variety of integrals (e.g., Steel et al. 1993). For the divergences considered here, a Hadamard conjugation allowing for a proportion of invariant sites and a gamma distribution of the substitution rates of sites free to vary, offers good accuracy (Waddell and Penny 1996, Waddell et al. 2007). If this is deemed too slow (typically greater than ~20 taxa with 2-state models), the fast approximations of Steel and Waddell (1998) may be used instead.

The program MS (Hudson 2002) was also used to test the predictions of our Excel spreadsheet calculations and to model the expected infinite sites spectrum. In turn, MS was run by a PERL script we wrote. This called MS, feed in the parameters for a hierarchically structured model of populations, with a single haplotype sampled from each population. Usually, 10 million independent samples of gene trees were taken and stored, then our script processed each tree, adding each internal edge length encountered to a table of all possible bipartitions. The edge parameters of this model are time measured in coalescent units, that is, the number of generations divided by the effective population size of the genes. We call these "g" units.

## 3 Results

The results below begin with a preliminary exploration structure in the data using up to date and appropriate methods of phylogenetic analysis, including residual resampling and NeighborNets. The relationship between the weighted site pattern vectors and tree space is discussed, particularly as the likelihood function is closely approximated by a weighted Euclidean space in the form of a $X^2$ statistic. Following this, the site pattern spectrum is intensively examined for various types of homogeneity, culminating in visualization of structure between chromosomes with Correspondence Analysis tweaked to reveal what seems to be the most interesting details, including potential details of male dominated gene infusion between Denisovan and Papuan lineages (the so called "Ron Jeremy Hypothesis"). Following this we fit a single "species" tree to the data using a variety of fit criteria with tweaks, and explore the stability of the key coalescent parameters. We then interrogate the data with a variety of sets of linear invariants on both the full spectrum, and on subsets of it (using the $P_1$ and $P_2$ statistics of Waddell et al. 2000 and 2001). These highlight different aspects of misfit to a single hierarchical coalescent model without reticulation. We then describe a coalescent model with reticulation (hybridization) between specified lineages and model a single mixing of Neanderthals with the ancestors of the out of Africa modern people. The fit of the predicted sequence spectrum is examined in detail. Following this we explore the model of a further mixing event between the Denisovan and the Papuan lineage, using MS. We then move from an infinite to a finite sites reticulate coalescent model with mutation and / or sequencing/processing errors using Excel spread sheet calculations. This brings a major improvement in fit of data to model. The fit the X chromosome data to the models generated by the autosomal data is checked, and the implied effective population size of X compared to other chromosomes to test the hypothesis that male effective population size is smaller than that of females (the so called "Genghis Khan Hypothesis"). Finally, we discuss how sub spectral methods, such as P, D and E statistics, are treated in a finite sites model framework.

### 3.1 Preliminary exploration

In this section we show a few useful phylogenetic methods for exploring inter population site spectrum data, including some techniques we have developed in the past few years. These are methods are not sufficient in themselves, but they are more advanced than those used in Reich et al. (2010) for preliminary exploration, thus it is useful to demonstrate them to readers.

Confronted with data like this there is a strong tendency to want to look at a tree. In particular, since the data represents many predominantly unlinked sites with incomplete coalescence, it is wise to use a least squares distance method (e.g., Bryant and Waddell 1998) since distances can remain additive in expectation on the species tree under a neutral infinite sites model (we confirm this empirically by running simulations with MS given the optimal parameter values found in later sections). Also, since the data are transversions only and the greatest distance to chimp is only 0.003 substitutions per site, it seems appropriate to work with an uncorrected Hamming or p-distance (sometimes called π in population genetics as opposed to phylogenetics, Swofford et al. 1996). In terms of a tree selection criterion, a flexi-weighted least squares model with weights proportional to the observed distance raised to power *P*, should give a much higher likelihood of the distance data than a BME or NJ tree (Waddell et al. 2011). In theory, the optimal weight for *P* with this data should be very close to 1 since the variance of a scaled binomially distributed distance like the proportional Hamming distance, p, is approximately proportional to p, when p is small (here all p-distances are less than 0.01) and if the only source of error is the multinomial (although the coalescent process does add some over-dispersion when there is any linkage).

For the data of Reich et al. (2010), the minimum value of $P$ by the g%SD criterion (which is monotonic with likelihood, Waddell et al. 2007, Waddell and Azad 2009) is found at $P = 0.2$, with a value of 0.235. At $P = 1$, the fit is slightly worse at g%SD = 0.251. Since a value of $P = 1$ is supported *a priori*, we will work with that. The resulting tree from PAUP* is shown in figure 1a. It is identical in topology to the NJ tree of Reich et al. (2010) with similar edge lengths. Applying residual resampling procedures (Waddell and Azad 2009, Waddell et al. 2011) to the distance data, all edges receive support values of 100%, except for the grouping of Han with Papuan, which receives support of 85%.

While a g%SD value of 0.251 seems very small, it is important to note that the external edges hugely dominate the data, such that the internal edges are just 2.4% of the total. This can make data with poorly resolves any internal edges seem very tree like by the g%SD statistic (Waddell et al. 2007). If the g%SD is scaled in proportion to just the internal edge weights, then the ig%SD rises to 10.3% (or 14.1% in the reduced bias form, Waddell et al. 2011), taking into account the number of fitted parameters. This is not nearly as tree like as would be expected for so many sites scattered across the genome.

Another potentially useful way to view this data is with NeighborNet (Bryant and Moulton 2004) combined with residual resampling (Waddell et al. 2011). The resulting labeled NeighborNet is shown in figure 1b. The data is, at a first glance fairly tree-like, although there are alternative splits showing up. The fit to a planner network is g%SD = 0.0769, which corrected for fitted parameters increases to 0.166%. Further, the sum of internal edge weights is 2.6% of the total weight, so ig%SD taking into account fitted parameters, is equal to 6.32%. This is a considerable improvement over a tree. This improvement in fit can also be queried using information statistics (Waddell et al. 2011). The planner NeighborNet graph beats the bifurcating tree by 47.2 AIC units and by 35.2 BIC units. However, since the number of parameters (splits = 22) in the NeighborNet is approaching the number of distances, 28, the AICc measure severely penalizes it and it ends up 129.2 AICc units worse than the bifurcating tree. However, if we were to drop either the NeighborNet edges with less then 50% residual resampling support (figure 1d) or filter all edges with a length of less than 0.0003 (figure 1d, except that the split DNP is retained in place of YFH), then the NeighborNet comes back to about equal with the tree in AICc units. Thus, unlike the example in Waddell et al. (2010) comparing the NeighborNet and tree of genetic distances to trace the origins of Jewish populations, a NeighborNet can be favored over a tree.

Equally important, and unlike Waddell et al. (2010), in the face of the competitive nature of residual resampling (which is necessary with such complex model selection), many of the NeighborNet splits have greater than 50% support (figure 1b and c). The most strongly supported split that is not compatible with the tree is the grouping of Han + French with 88.5% support at $P = 1$. This grouping is represented by a relatively short alternative split, with weight (length) of 0.0017. By contrast, the weightiest internal split separating modern humans from all others has length 0.014, while the external edges to human sequences are around 0.11 (based on the scaled distance matrix in the appendix of Reich et al. 2010, where the Chimp-San distance is 1). This split may be real as, to date, it is unclear exactly how to reconstruct the deeper genetic origins of Chinese (something we will consider further below).

Other well supported splits that conflict with the tree are Papuan with Chimp + Neanderthal + Denisova having 83.2% support (weight 0.0010) and a split of Denisova with Chimp having 87.3% (0.0010) (figure 1b and 1c). While it is unclear exactly what is causing these, one interpretation might be that Denisova is harboring some more "primitive" alleles than Neanderthal. That would be possible if the Denisovans had incorporated alleles from an even earlier out of Africa radiation, such as *Homo erectus*. The antiquity of some Papuan alleles in particular seems to be suggested by the split with Chimp + Denisova + Neanderthal. Aspects of this split are suggested by mixing of Papuans with Neanderthals (something shared with other non-African modern humans) and with a Denisovan relative (Reich et al. 2010). However, the Chimp being included in the split seems new and may also suggest the introgression of *Homo erectus*-like alleles particularly into Papuans. One interesting aspect of these splits is that they are both more than five times the length of the Han + French split (at ~0.0017 versus 0.0010)

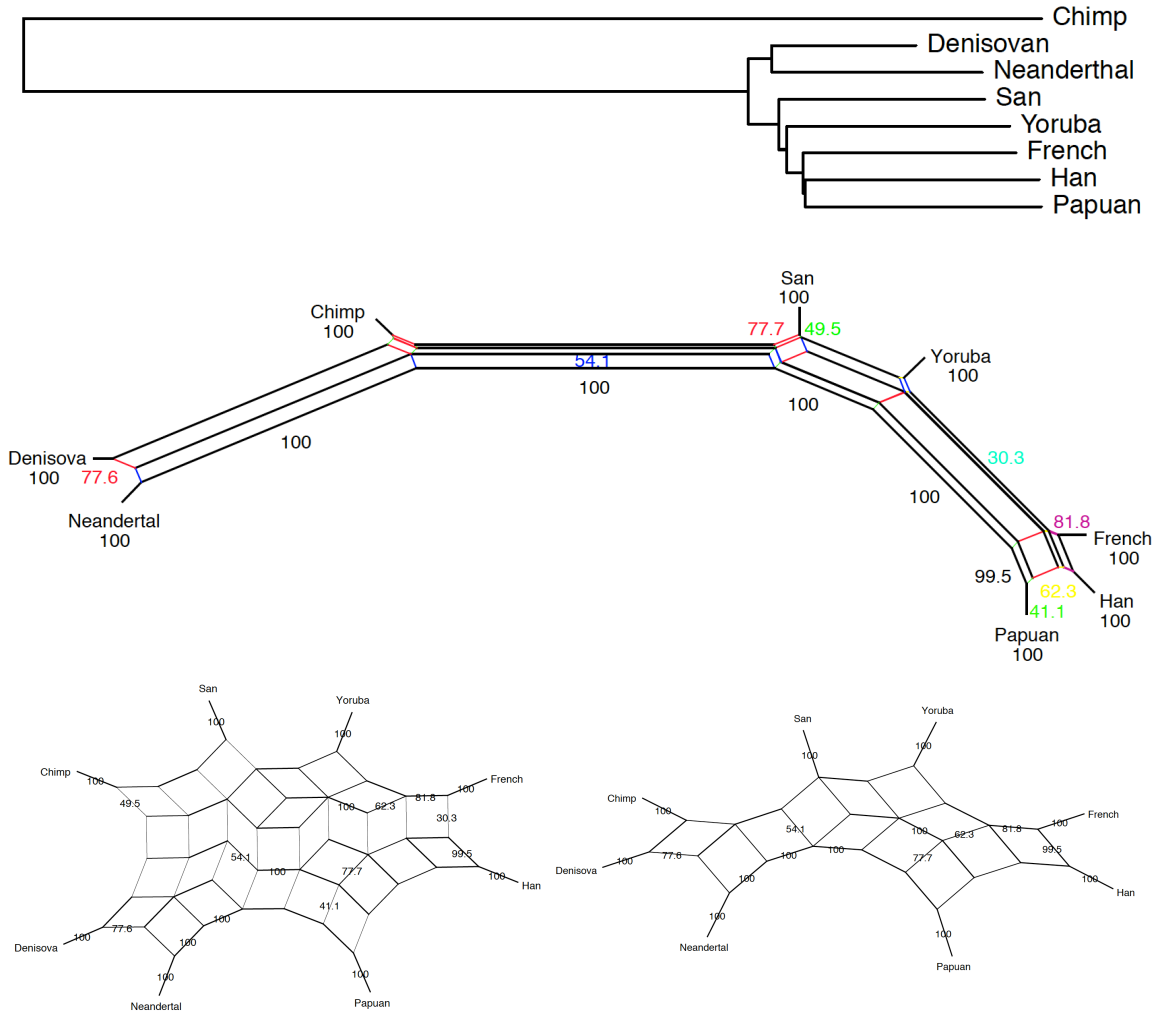

Figure 1. Fitting flexi-weighted least squares trees and NeighborNets to the site pattern frequency spectrum. (a) The flexi weighted least squares tree based on Hamming p-distances, with *P* = 1. All edges have residual resampling support of 100% except for the Han + Papuan split, which has support of 85%. The g%SD value (uncorrected for the fitted parameters) is 0.251% and corrected for fitted parameters is 0.336%. (b) The weighted NeighborNet with resampling support values (shown) estimated with *P* = 0. The g%SD value is 0.0769% and 0.154% corrected. All external edges except for chimp have been shrunk by a factor of 100, while the edge to chimp is shrunk by a factor of 1000 to improve viewability (the unscaled external edges are very similar to those of the tree in (a)). (c) The unweighted NeighborNet (YF split filtered out) and (d) The unweighted NeighborNet filtered to show only edges with greater than 50% residual resampling support. Switching to *P* = 1, yields the same network, but residual support values generally increase, that is HP:99.7%, HF:88.5, (CD) = NSYFHP:87.3, SYFH:83.2, NFHP:64.2, YFH:63.8, DNP:48.1, (CS) = DNYFHP:46.6, YF:39 and DNPH:34.7, with all other edge support staying at 100.

Other splits are not so well supported, but do appear in more than 50% of replicates. These include a narrow (0.00013) split of Han + French + Yoruba from all others, with 63.8% support. Whether this might reflect migration of the last 30,000 years amongst the more accessible parts of the world is unclear. The split of Neanderthal with all the out of African moderns has a length of 0.0005 and support of 64.2%, and at least in its existence, is consistent with Reich et al.'s (2010) analyses.

Finally, there appear a number of splits with between 40 and 50% support. One of these is the split of Neanderthal + Denisova + Papuan, that would be anticipated from the work of Reich et al. (2010) with a length of 0.00035 and support of 48.1%. There is also a split of San + Chimp with 46.6% support but a very low weight of 0.0001. A split of DNHP receives support of 34.7% and a low weight of 0.00017. Finally, there is a very weak split of French + Yoruba with a length of 0.00004 and support of 39%. Such a weak split is possibly below the level of sequencing/assembly error in this data. The San + Chimp split is also, to a fair extent, compatible with sequencing error due to its simple pairwise structure and low weight.

Overall, one of the most surprising features of the planner NeighborNet model is that it does not reverse the positions of Neanderthal and Denisova, so that Papuan could have a unique split with the Denisovan (as Reich et al. 2010 suggest the Papuan lineage received ~5% of its genes from that lineage). As we will see later, the apparent reason for this would seem to be that the distance from Denisova to Chimp is more strongly underestimated than that from Denisova to Papuan. The underestimation of the Denisova to Chimp distance could be due to Denisova harboring some very archaic alleles, or it could be sequencing error. That is, it might be direct convergence of Denisova to Chimp due to sequencing errors on the Denisova lineage. However, Denisova is reputed to have a particularly low sequencing error rate (Reich et al. 2010). It might alternatively be all the other genomes having a much higher sequencing error rate than Denisova, although if these are occurring independently and at a low rate, their main effect should be to simply extend the terminal edge length to each of these taxa.

All in all, the NeighborNet analysis clearly suggests that there is considerable structure in the residuals of the data fitted to a tree, that is clearly depict able with a planner graph. This is not to say that the remaining residual is negligible, but it does clearly suggest that more than a single species tree. This applies even to a distance based "species" tree, which can, in theory and simulations (unpublished data) adequately represent a hierarchically structured coalescent model, from which these data would *a priori*, be expected to come from.

**3.2 The likelihood surface of multiple phylogenetic/population genetic spectra**

The counts of the different types of site substitution pattern (the **f** vectors) define a proportional sequence site pattern (**s**) vector for each chromosome (Waddell et al. 1994). If each site pattern is sampled independently, then the **f** vector for each chromosome is a multinomially distributed, while the **s** vector has a scaled multinomial distribution. In turn, if these vectors are lined up against each other and indexed by the region they come from, then they jointly have a multinomial distribution when sampled independently. Under the infinite sites model, there is a natural and perfect correspondence between the bipartition patterns and the spectrum of the potential edges of all possible trees. Thus, the likelihood surface of the collective tree edge space (or more generally, the $\gamma$ vector, Hendy and Penny 1993) is that of the **s** vector. Deviations in this space may be measured with a likelihood ratio statistic (such as $G^2$ or the Kulback-Leibler distance measure) or it can be closely approximated with the $X^2$ deviance distance statistic (which has the advantage of intuitive interpretability since each cell has an asymptotic $\chi_1^2$ distribution under standard assumptions).

As we move away from the infinite sites model due to an increasing probability of there being more than one change of character state per site, there is a gradual break down of this approximation of the likelihood surface. The rate at which this approximation breaks down can be much reduced if a Hadamard conjugation (Hendy and Penny 1993) is performed (Waddell 1995), or, with more speed but less accuracy, conjugation approximations of Steel and Waddell (1998) may be used. The exact relationship requires the integration over continuously variable (gene tree edge lengths and site rates) and discontinuous (tree topology and coalescent history) space, which is currently not practical in any exact form. Of course, approximate numerical integration with methods such as Markov Chain Monte Carlo is also available, but often preclude a direct intuitive understanding of what is going on. However, these too can break down when the likelihood function itself is too complicated to estimate, necessitating approximate Bayesian chain or ABC type approaches.

Comparison of the **s** vectors for different chromosomes with a $X^2$ or KL statistic is a useful metric for detecting anomalies from the assumed i.i.d model (Waddell 1995). To facilitate this, it is useful to visualize the resulting full pairwise "distance" matrices between chromosomes with clustering or with multi-dimensional scaling. Preferably, both are done with the same measures of fit so that the results may be directly compared in a likelihood, AIC or BIC framework (Waddell et al. 2010).

By good fortune, a $X^2$ space (which may be thought of as a weighted Euclidean space) can also be analyzed with Correspondence Analysis (e.g., Greenacre 2007). This allows the interrelationship of the rows and the columns of such a space to be explored and contrasted with i.i.d assumptions. Here we use all these approaches to explore this interesting data set, to assess previous hypotheses and consider novel ones, and as important preparation for more explicit phylogenetic/population genetic models.

**3.3 Model free exploration of the data with clustering**

First calculated is the pairwise $X^2$ contingency-table statistic (or distance) between all pairs of $\hat{\mathbf{s}}$ chromosomal vectors for all the site pattern vectors from Reich et al. (2010). Asymptotically (as the number of site patterns observed or the sequence length $c \rightarrow \infty$), and given i.i.d. assumptions, if the two vectors are samples from the same true parameters, the $X^2$ statistic should converge to a $\chi^2$ distribution with degrees of freedom equal to the number of site patterns, minus the number of estimated parameters. In this case, we start by considering all possible bipartitions, therefore our number of site patterns is $2^{t-1}$ (where $t$ is the number of sequences being considered) per sequence, which in this case is $2^7 - 1 = 127$ independent patterns so there are 254 patterns in total (since two sequences are being compared). The number of estimated parameters is the row averages (less one) of which there are also 127. Thus the degrees of freedom are equal to 127, or more generally $n - 1$, where $n$ is the number of patterns. This is effectively a two-vector test of homogeneity (e.g., Agresti 1990).

With this type of data it is important to examine carefully a number of subdivisions of the data. In terms of sequence site patterns, two prominent subdivisions are to first ignore the constant sites and secondly to ignore the singletons as well, leaving just the parsimony informative sites (e.g., Hendy and Penny 1993). Ignoring the constant site patterns adjusts for a potentially powerful scalar effect of the overall rate of evolution on a chromosome. Ignoring the singletons, or site patterns with just one sequence different to all others leaves just the bipartitions corresponding to all possible internal edges in the tree or network linking the sequences. The singleton counts are particularly susceptible to sequencing errors. In this data, the true signal is of a similar magnitude to the sequencing error signal according to the supplementary material of Reich et al. (2010). Here it is reported that a transversion per site occurs at a frequency of ~$10^{-4}$ in the human sequences, which is roughly the same size sequencing error rate in many of these sequences. Independent sequencing errors have a proportionately greatest impact on singletons, followed by the simplest informative site patterns, such as two different from all the others, and less effect on complex patterns such as four different to all the others. In terms of the genomic blocks, we treat the random blocks differently to the remainder of their chromosome, since they are more poorly localized. In addition, Chromosome X is typically found in two copies in a female, but only one in a male, which gives it an effective population size of only 75% that of the autosomes if the effective population size of males and females is equal.

It is also important to bear in mind the sample size when assessing convergence to asymptotic conditions for the $X^2$ statistic or distance. A general rule of thumb is that as long as 75% of all expected entries in a comparison of **f** vectors are equal to or greater than 5, and no expected values are less than 1, a reasonable approximation holds. To facilitate use of this rule, we report the expected values in each comparison.

An example of visualizing the clustering of site pattern vectors is given in figure 2 using NeighborNet and NJ. Notice that the general impression is very similar. The outliers include the identity pattern, the singleton to chimp (DNYSFHP), and a selection of patterns that may be linked to relatively recent inter breeding. These include NP, DNP, DP, and NH. The pattern NYSFHP is also particularly uneven across autosomes, which is interesting since it is the pattern that would most obviously mark gene flow from *H. erectus* to the *H. denisova* lineage, which is not so recent. When the X chromosome is added in, the same patterns appear as outliers, while removing constant sites and singletons, the outliers remain the informative patterns mentioned earlier in this paragraph (with or without X included). When fitting an least squares tree with $P = 0$ the residual error reported was surprisingly large, which cautions us corroborate the observations above in other ways.

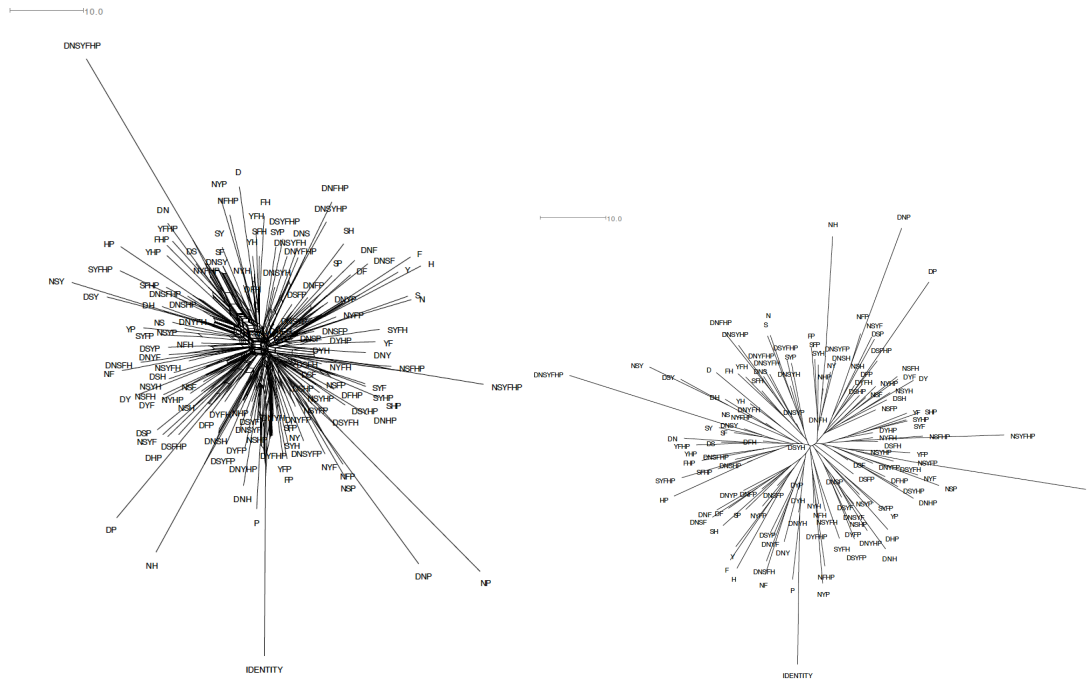

Figure 2. Visualizing the heterogeneity of site patterns based on the pairwise $X^2$ distance between their frequencies across autosome chromosomes. (a) Using NeighborNet in SplitsTree4 with a reported fit of 0.8035 (b) Using Neighbor Joining with a reported fit of 0.8092. Note, the fit reported by splits tree is the percentage of all variance in the data explained (see Waddell et al. 2007 describing how this related to %SD measures). The flexi-weighted least squares tree with $P = 0$ tree is broadly similar but reports and a g%SD fit of 87.4% which is not terribly good. The expected $X^2$ asymptotic distance between a pair of site patterns if there are no differences in their true frequencies across chromosomes is 22 in this case. The scale bar indicates a distance of 10 units. In the electronic version it is possible to zoom in several times to view these plots in detail.

A useful feature of the pairwise distances between site pattern vectors is that objects for which the distance is being measured can be included or excluded from the visualization analysis simply by removing them. With other diagnostics used herein, such as deviation from the global average (equals the expected values under independence), then the analysis typically needs to be rerun from the exact initial data matrix (for example, with correspondence analysis).

**3.4 Gauging the effect of linkage**

Another useful feature of the $X^2$ statistic is its relatively simple asymptotic relationship to the number of independent objects being sampled. For example, if we were to halve the effective number of independent random samples by automatically choosing two of an object (rather than 1), then the variance of the observed count per cell would basically double. If this logic is coupled with the biological knowledge that, no matter how deviant the mating behavior, the assortment of chromosomes from the mother and the father is random, then it allows a direct estimate of the scale of variance increase due to linkage. Also bundled with this variance will be presently unmodelable factors such as specific instances of selection, so these statistics will also give an upper bound estimate of the potential size of these factors on overall fit. Such an estimate on the expected precision measured by fit statistics is very useful when used in a quasi-likelihood framework.

The visualization in figure 3a is based on pairwise $X^2$ distance between rows (chromosomes) of the original joint table. It can be seen that chromosome X and chromosome 16 appear to the two most marked outliers. Much of this is due to the effect of constant sites, which are heavily influenced by the average rate of substitution. Removing this site pattern, chromosome X appears to be the only marked outlier (result not shown). Finally, after removing the singleton site patterns as well, figure 3b shows the results based on just the parsimony informative site patterns. The trees in this figure are not to scale and the average distance between chromosomes is hugely reduced. Here we see that chromosome X appears the only real outlier and it is not nearly as pronounced as previously. In addition, the fit of the distances to a tree is far better with a g%SD of 8.6%.

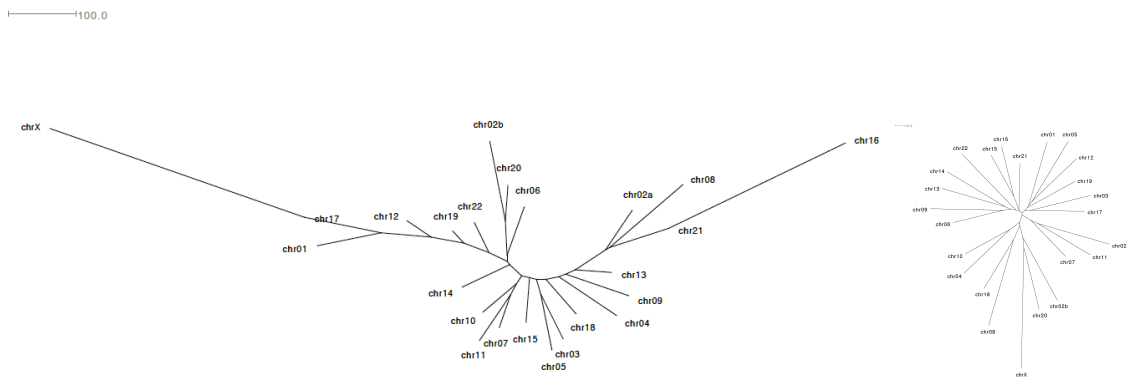

Figure 3. Visualizing the heterogeneity of chromosomes based on the $X^2$ distance between their pairwise frequencies across site patterns. (a) Using all site patterns the NJ tree has a reported fit of 0.8348 (b) Using just the informative site patterns the fit improves to 0.9346. When fitting with least squares $P = 0$, the tree (not shown) is nearly identical to the NJ tree of figure 3b and the average geometric percent standard deviation has dropped to 8.6%.

Overall, the tree of figure 2b is not that far from a star tree. In addition, it is useful to note that the two arms of chromosome 2 (which are two separate chromosomes in the chimp) are not particularly close to each other, suggesting they may be functional as independent as most chromosomes. If we take all the random blocks and add them together they approach the 75% rule for asymptotic approximation discussed below. When they are added into the visualization shown in figure 2 they appear very normal, that is, they are near the midpoint of the tree without a long external edge. However, later, we see that when analyzed with correspondence analysis, while they have a very low inertia, they do appear to have high inertia on the first two axes, and appear deficient in patterns close to the species tree but enriched in patterns far from either the species tree or probable major introgressions. This might be a sign that they are derived from an increased proportion of reads that do not correctly align to their homologues and we generally discard them from analyses (they comprise less than 1% of all the data).

Figure 4 shows the result of measuring the fluctuations of informative pattern frequencies with the $X^2$ distance of each chromosome against their global average across all chromosomes (which is the same as the expected values based on a global multinomial i.i.d. model). This is the same as looking at the marginal column sums of the $X^2$ statistic in an *n* x *m* contingency table test of independence, where columns are chromosomes and rows are the site patterns. In the language of Correspondence Analysis, considered below, these are the total inertias of each column of the original data table (which are the $X^2$ column sums renormalized to sum to 1000).

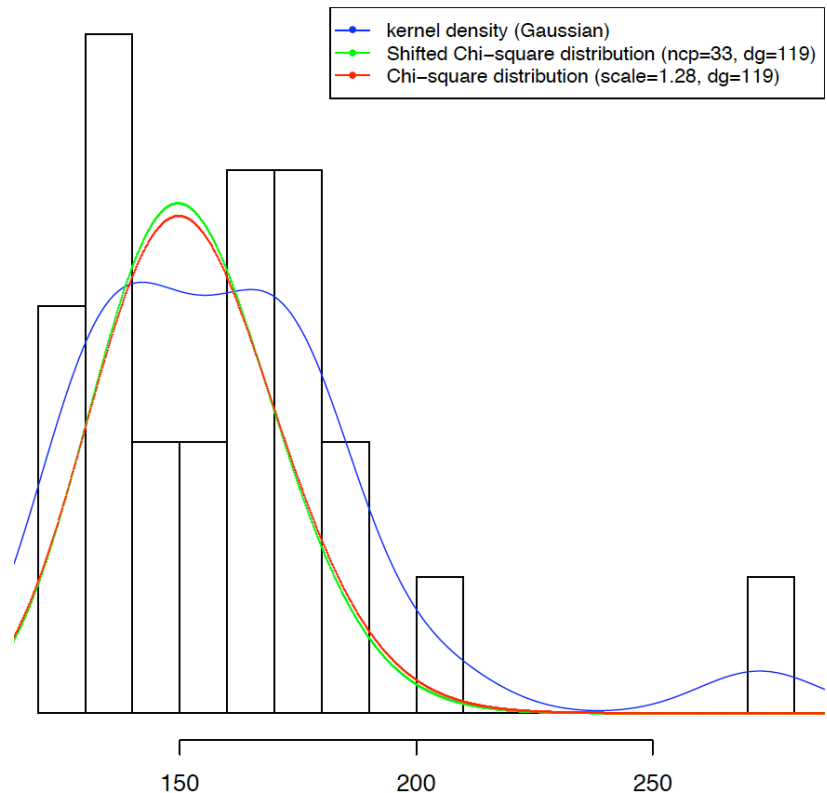

Figure 4. A histogram of the $X^2$ distance of each chromosome to the global average of site pattern frequencies weighted assuming independence (proportional to inertia). The blue line is the result of fitting a smoothed curve based on a normal distribution kernel density, while the green and red lines are the shapes of a non-central and a scaled chi-square distribution, respectively. Except for the outlying chromosome X that is at the far right, the chi-square distributions offer a modest approximation to what is observed. If site patterns were i.i.d. across chromosomes then the expected value would be total site patterns (128) minus constant columns (1) minus singletons (8) minus 1 degree of freedom = 118. The actual mean (excluding X) is closer to 155, so ~ 30% bigger than expected under independence. The ordering of chromosomes from left to right is 7, 21, 10, 17, 6, 1, 15, 19, 12, 16, 4, 3, 13, 18, 2b, 5, 11, 14, 22, 20, 9, 2a, 8 and X. Removing chromosome X makes almost no difference to the rank or relative inertia of the autosomes, as might be expected as X is a low mass or count chromosome (here mass refers to the number of informative site patterns recorded from it, not its physical size).

It is also possible to look at the row sums, which record the fluctuation of the frequency of each informative site pattern across chromosomes. Figure 5 shows such plots. There are clearly a number of outliers, and these are the same ones identified in the pairwise statistics of the previous section. That is what we will call the Denisova/*H. erectus* pattern, plus the NP, DNP, NH, DP patterns. All these patterns would seem to be associated with potential introgression events. In that regard, it is interesting that DNP appears as high on the list as DP, perhaps due to the Denisovan lineage that introgressed into Papuans being a distant relative of the sequenced Denisovan and not being particularly coalesced for the patterns found in that individual. These are followed by the HP pattern, which is intriguing, as this may be due to the recent (past few thousand years) introgression of Austronesian genes from East Asia into Papuans, or simply due to this being the last split in the species tree. It is intriguing that these patterns retain their inertial ranking with or without chromosome X included. This may be a reflection of their overall higher standard deviation due to being generated by relatively recent events (which, with the exception of the D / *H. erectus* pattern, would all seem to be less than 60,000 years before present) while the coalescences generating most other patterns are much older on average and have had much more time to unlink from each other. Other patterns logically associated with the hypothesized Neanderthal / out of Africa population, which include NF, NFP, NFHP, coincide with what may be a second mode in the distribution of figure 5. It is unclear why the other species tree patterns, including SYFHP, YFHP, DN, and finally, FHP, all appear high in the inertia list (with or without X included), although these are predominantly generated by more recent coalescences, and hence it may be an expected linkage effect increasing their variance across chromosomes.

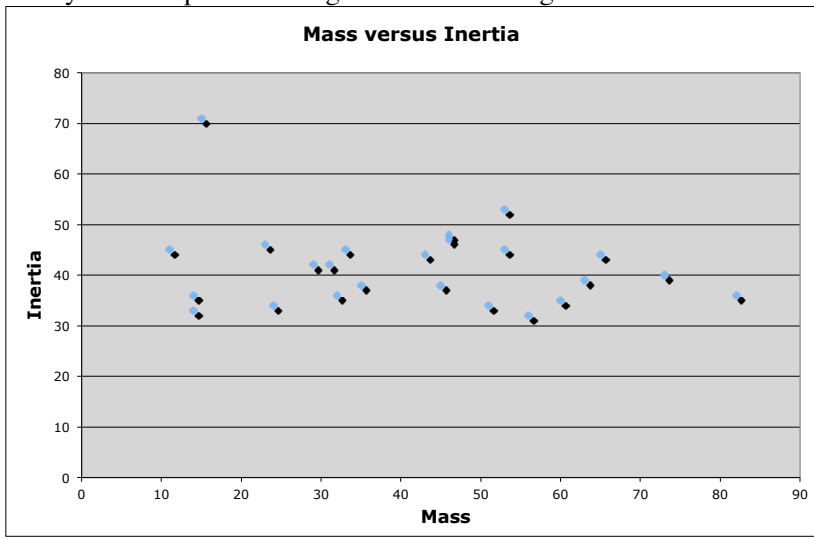

Figure 5. A plot of mass versus inertia for the chromosomes. All the chromosomes except 19, 21, 22 and X easily meet the 75% greater than or equal to 5 and none less than 1 expected value rule for approximating the asymptotic properties of the $X^2$ statistic.
Chromosome 19 has 26% of values below 5, but not less than 1, while 21 has 29%, 22 has 39 %, and X has 24%. Only X is an outlier.

We have mentioned the D / *H. erectus* pattern (NSYFHP) as showing high deviance and inertia. There are two other patterns that might readily reflect an earlier out of Africa mixing with even more archaic hominids such as *H. erectus*. These are N / *H. erectus* (DSYFHP) and DN / *H. erectus* (SYFHP). The last one is also a species tree pattern, while the former two are only 1 coalescent event shy of it. It is interesting to note that two of these patterns do show very high inertia (NSYFHP and SYFHP) while the third one is in the middle of pack in this regard. Also confusing this interpretation at present is the fact that while Denisova has a sequencing error rate that appears very low (as low or lower than the modern human sequences used according to tables in the supplementary material of Reich et al. 2010), Neanderthal sequence has spotty coverage and an apparently high sequencing error rate. Neanderthal flickering off and on, respectively, in a background of common DNSYFHP and SYFHP patterns could be a cause. However, other patterns readily generated in this way, such as NY or NS do not show high inertia at all, in fact they are below average).

Overall, then, the constant site pattern and the singletons have very high mass and high inertia, patterns related to the introgressions proposed by Reich et al. (2010) generally have low mass and high inertia, while patterns on the species tree tend to have intermediate mass and often high inertia. However, in terms of inertia per unit of mass, doubling the mass should increase the inertia by a factor of two, tripling the mass by a factor of three and so on. On this scale the singletons have low inertia per unit mass. The patterns related to hypothesized introgression events tend to have very high inertia per unit of mass, while the patterns coinciding with the species tree are intermediate.

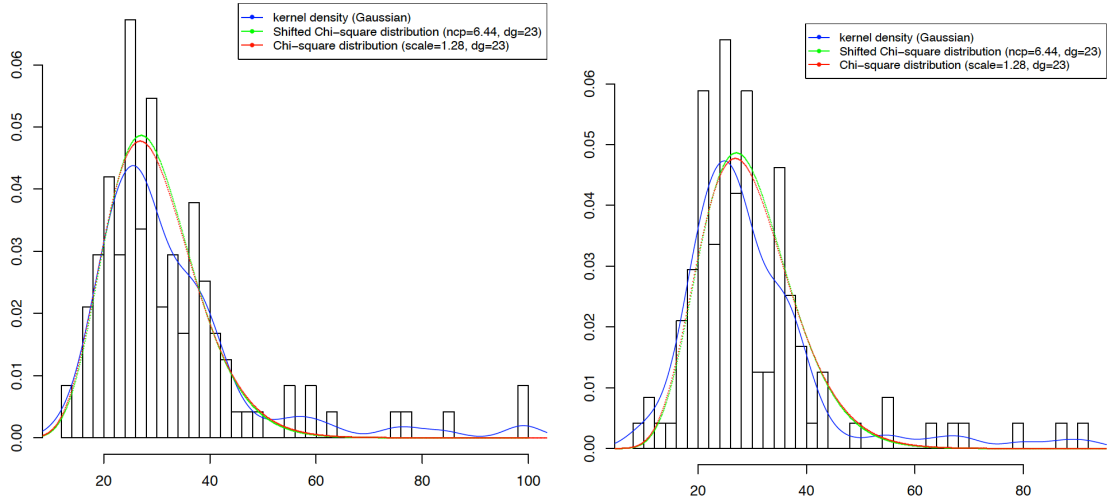

Figure 6(a). Histogram of the $X^2$ statistic for each site pattern on each chromosome against the global frequency (proportional to the inertia of rows in this case) for all chromosomes including X. (b) As in (a) but excluding chromosome X. The blue line is the result of fitting a smoothed curve based on a normal distribution kernel density, while the green and red lines are the shapes of a non-central and a scaled chi-square distribution, respectively. If site patterns were i.i.d. across chromosomes then the expected value would be 23 in (a) and 22 in (b). The actual mean values are ~30% bigger than expected under independence. The ordering of site patterns from left to right is DSYH, DNSYP, DFH, DYP, DNFH, YH, SF, NYFHP, NSYHP, DYH, DSF, DSFP, DNYH, DNYFH, DNSHP, DNSFP, DNSFHP, DNSY, DNSYF, SP, SFH, NHP, NFH, NYH, NYFH, NS, NSF, NSYFP, DFHP, DYHP, DYFHP, DSHP, DSFH, DSYF, DNFP, DNYP, DNYFP, DNS, DNSP, DNSYH, SFHP, SY, SYH, NY, NYHP, NYFP, NSH, NSHP, NSFP, NSYP, NSYFH, DH, DFP, DYFP, DYFH, DS, DSYP, DSYFHP, DNYHP, DNYFHP, DNSH, DNSYFP, DNSYFH, FP, YP, YF, YFH, SFP, SYHP, NSFH, NSYH, DSH, DSYFP, DSYFH, DNHP, DNY, YHP, SYP, SYFP, NYF, NSYF, DHP, DSFHP, DSYHP, DNYF, DNSFH, FHP, SH, SHP, SYFH, NF, NFP, NFHP, NYP, NSFHP, DF, DY, DYF, DNF, FH, YFP, NSP, DSP, DSY, DNH, DNFHP, DNSYHP, YFHP, NSY, DN, DNSF, SYF, SYFHP, HP, DP, NH, DNP, NP, and finally, NSYFHP. Removing chromosome X makes almost no difference to the rank or relative inertia of the site patterns; the most noticeable difference being that the "erectus" pattern now has second highest inertia across the chromosomes. The vectors used all meet the 75% rule for asymptotic convergence.

Table 1 shows a different way of assessing the impact of linkage and erratic selection on any overall fit statistic. Here all chromosomes are randomly assigned to two nearly equally sized sets. This half and half approach should provide a robust estimate the $X^2$ statistics mean and variance between the data and a model prediction. Note that the relative proportion of constant sites is implied to fluctuate between chromosomes, as is the proportion of singletons. The extent of this deviation in relation to the informative sites is exaggerated by the fact that only 0.043% of sites are informative, and only 0.491% of sites are non-constant.

Table 1. The $X^2$ statistic between randomly split sets of chromosomes (based on 10,000 replicates, with Chr2a and 2b fused). The first number in each cell is the mean value and the second the standard deviation.

|                  | Autosomes          | Autosomes + X      | Autosomes + Random |
|------------------|--------------------|--------------------|--------------------|
| All sites        | 334.2561 (171.0058)| 342.4778 (181.8389)| 333.0322 (170.8260)|
| Minus Identity   | 233.4396 (65.2967) | 238.5651 (67.9386) | 233.7334 (66.2594) |
| Informative Only | 163.4504 (26.0422) | 165.8860 (26.8971) | 162.8040 (26.4666) |

This approach to assessing inflated $X^2$ values should have some robustness in suggesting the cause is linkage and not something more localized. For example, if the deviation was caused by just one cell (that is, one type of pattern in one chromosome), then the chromosome with this cell should appear as an outlier in figure 4, while the site pattern should also appear as an outlier in the row sum plot of figure 6. Correspondence analysis, applied below, is ideal for visualizing such associations. The degree of inflation we infer here is around 1.35. This amount is neither excessively large nor small, but it does give us a good guide to how well we could hope to model this data with spectral species-tree coalescent methods that do not explicitly model recombination events (or any other method that ignores recombination events). In future it may be useful to further diagnose this overdispersion by looking in close detail at where specific patterns may be linked in the genome. Obviously, the extent of overdispersion could rise substantially with new data, as the present data set samples only 300 million sites or 146,019 informative sites, which will on average be approximately 25000 bases apart. If the amount of data rose 5 fold, then the average informative site would be only 5kb apart, which is much closer. How rapidly evidence of linkage rises by this gross statistic will depend on how factors such as recombination hotspots are scattered across the genome and how they have shifted through time in different populations, plus the age of the different haplotypes that specifically generate particular site patterns.

### 3.5 Correspondence analysis

The result of applying correspondence analysis to the autosomal plus X informative site patterns is show in figures 7 and 8. While the four axes in these two plots show only 20% then 34.7 % of the total inertia, the previous analyses suggest that random noise due to fairly low levels of linkage may explain much of the remaining deviations from an i.i.d. multinomial model. There may also should be some biological effect due to the lower effective population size of chromosome X, and this too can be checked for. However, it is interesting that the normalized inertia of the species tree patterns shows almost no change when the autosomes are considered alone without X (68 versus 67), suggesting that the smaller effective population size of X is not having a very strong effect on variance of patterns. Since X is an outlier, this suggests it may have more to do with its specific sex-linked properties

The first two axes are defined predominantly by chromosome X and 8, with lesser contributions by chromosomes 20 and 12. The species tree patterns show up prominently, but while one is drawn towards X (YFHP), with another two being drawn very slightly towards X (DN and FHP) and the remainder (SYFHP and HP) are being drawn more strongly towards chromosome 8 and they are decreased in X relative to the autosomal average. Another point is that the Neanderthal patterns for out of Africa are deficient on chromosome 8, while a Neanderthal plus SY pattern (not associated with any specific event as yet) is enriched on 8. Finally, the DP pattern and the D / *H. erectus* pattern are both deficient on X in particular. The decrease in frequency of the DP pattern on X, particularly when compared to the NP pattern (which is near autosomal average frequency on X) suggests the possibility of asymmetric gene flow in this introgression event. If so, it would seem that this might be most readily explained by greater survival and reproduction of the offspring of Denisova males impregnating the modern human female ancestors of Papuans rather than the other way around.

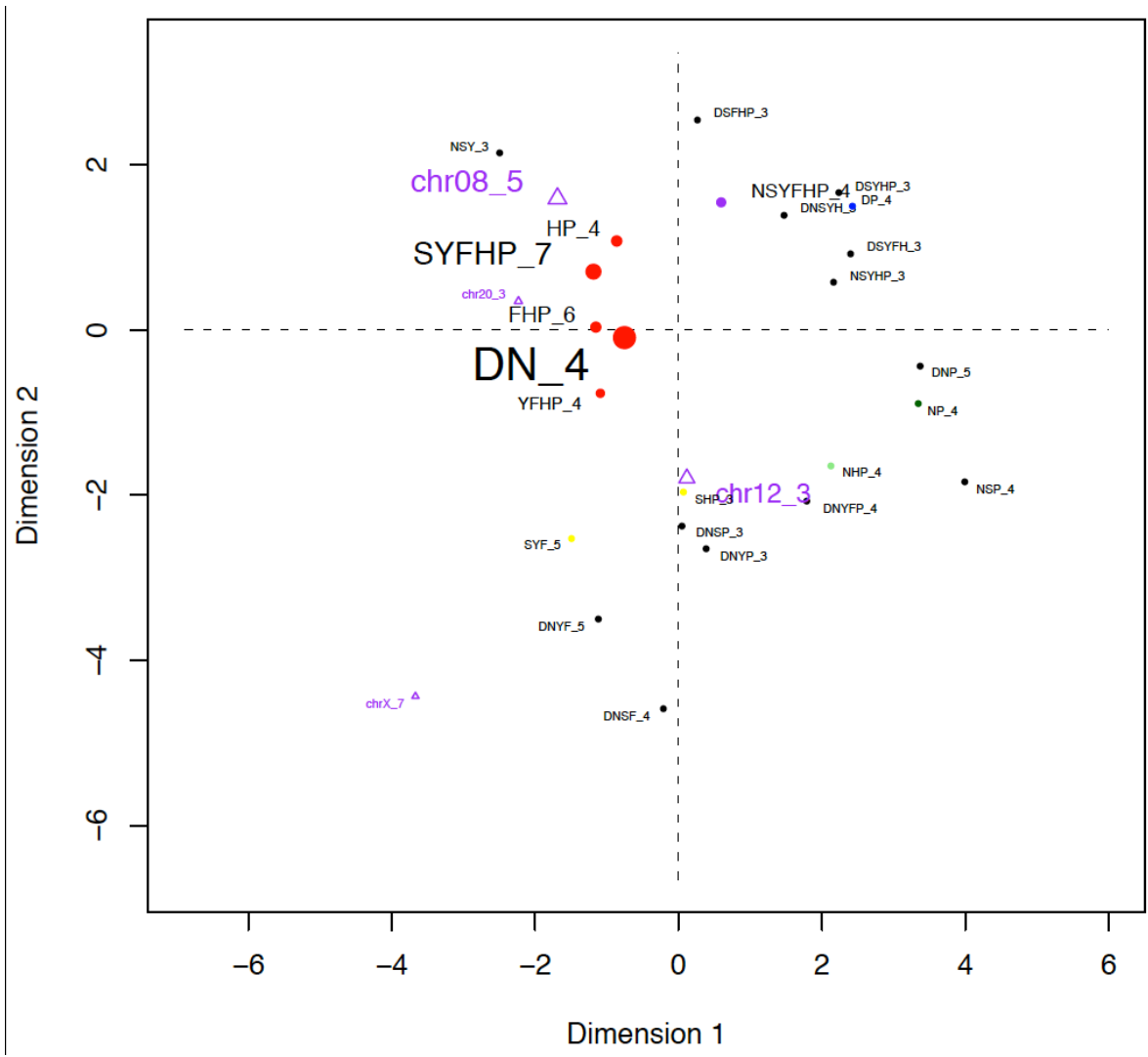

Figure 7. The first two dimensions of the correspondence analysis applied to the informative site patterns of all chromosomes (not including the random blocks). The first dimension explains 11.6% and the second 8.4% of the total variance (inertia). In these and subsequent plots the color coding red is for a species tree pattern, orange, a pattern that can be generated by missing one coalescence on the species tree, yellow, a pattern that can be generated by missing two coalescences on the species tree, dark green a pattern readily generated by introgression of out of Africa modern with a Neanderthal African (e.g., NP, NH, or NF), light green, as previously but shared by more than 1 modern, blue, a pattern readily explained by introgression of a member of the Denisova lineage with Papuans (DP and DNP), while violet might be leftovers of a previous out of Africa, which could be Denisova or Neanderthal ancestors meeting *H. erectus* (patterns NSYFHP or DSYFHP; the third pattern SYFHP is cryptic since it is also a species tree pattern). Chromosomes are marked by purple triangles. Labels and symbols are scaled according to their mass, while the name is followed by underscore then the quality of a point rounded to the nearest integer in the range of 0 to 10. Only objects with a quality of 3 or above (thus the diagram represents at least ¼ of their total inertia) are shown to reduce clutter and hopefully focus on the most salient results. Excess of the pattern NSYFHP over DSYFHP is consistent with a "Denisova / *H. erectus*" interbreeding. This hypothesis is supported to some degree by the mtDNA evidence (e.g. Krause et al. 2010, although somewhat dismissed by Durand and Slatkin in Reich et al. 2010). Further, the pattern DSYFHP is more readily generated by convergence of Neanderthal to Chimp via a sequencing error, by a factor of nearly 4 to 1 according to the error estimates in table S10.3 of Reich et al (2010). On the other hand, a pattern such as NSYFHP is more likely than DSYFHP due to Neanderthals mixing with out of Africa people (although the observed ratio is slightly larger at 1.17 that of NFHP to DFHP at 1.09). Diffusion of Neanderthal genes into Africa in addition, might also help to explain this imbalance. Inferring the relevance of such an outlying pattern, such as NSYFHP, requires detailed modeling of mixed hierarchical coalescent scenarios, which is attempted in later sections.

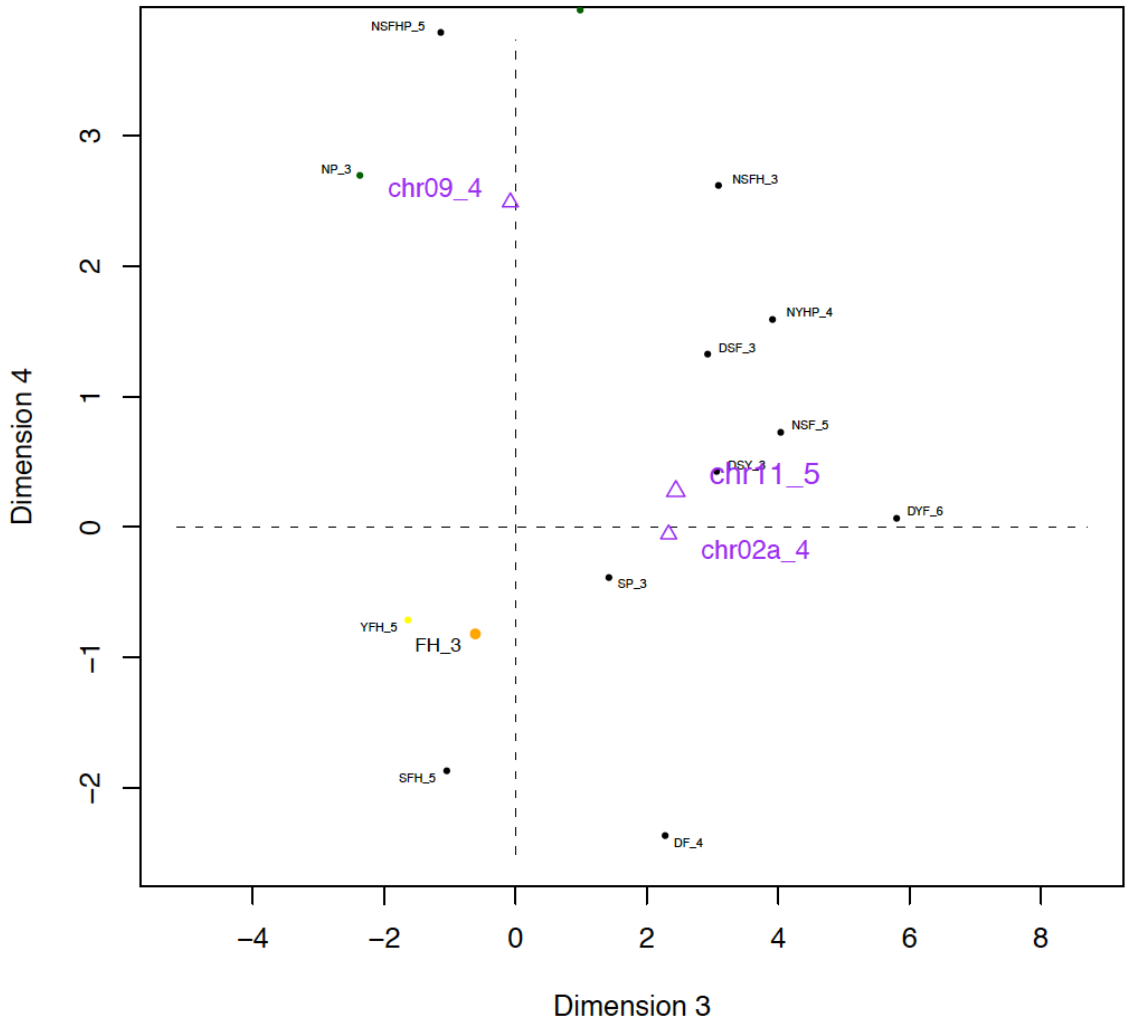

Figure 8. As for figure * but the third and fourth dimensions are shown, which, respectively, explain 7.7% and 7.0% of the total inertia.

While the origin of the unusual features of the NSYFHP pattern is just a hypothesis at this stage, it is testable and deserves a name, so we call it the "Ron Jeremy hypothesis" (after the accomplished American thespian Ron Jeremy, who is adroit at debauching modern young women, whose father's might well call him a Neanderthal or a Denisovan, and who looks remarkably like reconstructions of these archaic humans in museums, including being very big boned). Supporting the hypothesis that this was an asymmetrical mating event is the finding that the recombination rate on X is similar to that on other chromosomes, so its intrinsic variance should be no higher (Jensen-Seaman et al. 2004). Combined with the fact that the density of markers (site patterns) on X is considerably lower than that on the autosomes (at least partly due to its lower mutation rate), it seems probable that the general linkage disequilibrium between site patterns on X is lower than on most autosomes. Similarly, we may refer to the low frequency of the NSYFHP on the X chromosome as "Ron's Grandfather hypothesis" which is the mixing of the Denisovan lineage with an even more ancient hominid lineage due to a male biased infusion.

On the third and fourth axes (figure 8) there is no general theme of particular types of pattern emerging as there was in axes 1 and 2. There are a few scattered implications, such that the pattern NP is particularly enriched on chromosome 9, but nothing suggesting a general pattern.

### 3.6 Fitting the full spectrum to a species-tree with coalescence model

Here we expand on the type of parametric spectral fitting initiated in Waddell (1995) based on the calculations of Hudson (1992). The key point, as discussed elsewhere herein, is to calculate the total length of all bipartitions induced by all possible gene trees. More colloquially, in Hadamard parlance, it would be to calculate the expected edge length spectrum. Under the infinite sites model and assuming random mutations and recombination falling away with some fixed function of distance, the sequence spectrum is expected to converge to the edge length spectrum as $c$, the sequence length, goes to infinity.

In order to make these calculations it is necessary to find populations of gene trees for which we can calculate the expected edge lengths. To do this we break all gene trees down by the pattern of their coalescences, then by their shape, and, finally, if need be, by their history (that is, which coalescence happened first, second, etc.). For the case of a seven species rooted tree with the shape (C,((D,N),(S,(Y,(F,(H,P)))))), the problem was set to an second year undergraduate in physics as a research credit exercise (see Ramos and Waddell, arXiv article) and the results implemented into an Excel spread sheet. Optimization of parameter values used the robust numerical optimizer within Excel, which we have found to be generally reliable with these types of calculations in previous instances (e.g., Waddell 1995), and best checked by trying different random starting points. To check the accuracy of the results MS was set to the same parameters and run to produce millions of independent gene trees. Resulting internal edges were then gathered together with a PERL script and represented as a binary indexed vector following Hendy and Penny (1993).

The complexity of these calculations explode as a rather nasty exponential, so the complexity of the calculations is thousands of times larger than with a three or even four species tree. Comparing the results with MS revealed a series of mistakes in terms of which values were put into which cells of the spreadsheet. After refining these, there were still some differences, but on a percentage basis these are relatively small in all cases. To alleviate these errors we instituted an iterative local correction method.

The first choice to be made is which objective function to optimize (here minimize) between the observed and the expected informative site pattern frequencies. Two obvious choices are the $G^2$ likelihood ratio statistic (thereby making a maximum likelihood estimate of the free parameters, namely g1 to g5, which are time measured in generations divided by effective population size of the genes) or alternatively the $X^2$ statistic. As in Waddell (1995) we will explore both and expect both to give very similar answers. For $G^2$ the minimum achieved was 3231.48 with parameters g1 to g5 taking values of 0.028483, 0.173353, 0.094845, 0.348033 and 0.273664, respectively. For minimum $X^2$ the optimal solution was 0.026829, 0.171210, 0.092990, 0.342386, 0.266800, respectively for g1 to g5 to achieve an optimized $X^2$ statistic of 3473.97.

For the $X^2$ statistic we then ran MS for 10 million samples at these parameter settings. The resulting predicted spectrum was compared to the observed spectrum and the fit was indeed better, although not on all cells. Numerical error due to doing 10 million versus a larger number was small (later estimated to have a variance of about 6 units). At this point a local correction to the values being predicted by the spreadsheet was made (by adding the difference of the MS values to the predicted spectrum at the first set of optimal values). Doing this the fit improved from 3473.97 to 3175.02 (we denote this augmented $X^2$ statistic $X^{2'}$). Next, we reoptimized the free parameters to find the new optimum with the locally corrected predicted values. This reduced $X^{2'}$ to 3172.83 with parameter values 0.024947, 0.171766, 0.092427, 0.339571 and 0.266852. At these parameter values another MS simulation was run for 10 million samples and the spectrum again used to locally correct the values predicted by the spreadsheet. Reoptimizing the free parameter values according to the re-locally corrected spreadsheet values, we arrive at the solution with $X^{2''} = 3150.39$'' and parameter values of 0.025994, 0.171378, 0.092534, 0.339331 and 0.267358. Going one iteration further yields $X^{2'''}$ with an optimized fit of 3152.53 and parameter values 0.027933444, 0.175100731, 0.095006025, 0.344262573 and 0.27306159. At this point there is convergence in the fit value with the tolerance of 10 million random replicates from MS, but the parameters are still wandering somewhat. However, as we will see later, all estimates are within 2 s.e. on the final round of iteration of their previous cycles values.

To check whether there is a clear local attractor we started at a different set of parameter values. In this case, 0.1, 0.1, .1, 0.3 and 0.3 were used as the starting values and an MS prediction with 10 million replicates was made, then continued with the iteration sequence. The first cycle $X^2$ was 5077.75 and $X^{2'}$ was 4852.92. Optimizing with $X^{2'}$ improves the fit to 3254.93 with parameter values 0.024195, 0.172100, 0.091807, 0.339959 and 0.267993 for g1 to g5 (at this point the $X^2$ value was 3476.52). Updating with new MS predictions and optimizing with $X^{2''}$ yielded a minimum of 3130.76 with parameters of 0.0232710081, 0.1687159905, 0.0906498093, 0.3343395592 and 0.2612738259. Optimizing with $X^{2'''}$ in turn gave a solution of 3149.55 with values of 0.0227709286, 0.1689006031, 0.0903369873, 0.3342731036 and 0.2606716073. Some of these are about 2 s.e. from their expected values, and it is probably partly due to the MS correction factor having a s.e. itself of about 6 units (which with two different random correction factors suggests a s.e. of about 10 units between random starting points).

Doing the same thing for $G^2$ with correction the MS correction term based on parameter values of 0.1, 0.1, 0.1, 0.3, and 0.3 (for g1-g5) the optimized $G^{2'}$ was 2997.95 and parameter values were 0.026312, 0.174663, 0.093496, 0.345733 and 0.274174. Iterating again we arrive at $G^{2''}$ = 2900.82 and parameter values of 0.0295613399, 0.1774256720, 0.0957332632. 0.3510034118 and 0.2794670139. One more iteration with $G^{2'''}$ had a minimum of 2911.69 and parameter values of 0.0276910902, 0.1745604708, 0.0936033294, 0.3455494590 and 0.2728650303. These parameter values are all very close to those found previously with $G^2$ as the fit criterion.

As in Waddell and Azad (2009) it is useful to convert likelihoods into root mean square percent standard deviations between data and expected values to give them both a geometric interpretation, and an intuitive feel for the quality of the models predictions. For the $X^2$ statistic this is straightforward. In the case of the uniterated solution we have

$$\%SD[X^2] = \left( \frac{d.f.}{\sum_{i=1}^{N} expt_i} \right)^{0.5} \left( \frac{1}{d.f.} \sum_{i=1}^{N} \frac{(obs_i - expt_i)^2}{expt_i} \right) \times 100\%$$

Eq(1)

or 15.42% in the case of the first minimum $X^2$ model run above. The sum of percentage absolute deviations is similar at 14.83%.

If these data were perfectly multinomial we can calculate the expected deviation by noting that the second sum, under the model, is the expected value of the X^2 statistic, which in this case is 119 - $k$ ($k$ = 5 fitted parameters), so the whole second terms goes to 1. Thus, with 146,019 sites we would expect this number to be around 2.79%, while if they followed our prediction of the $X^2$ showing an inflation factor of around 1.5 then we would expect around 3.42%. Thus, worsening of the %SD by about 12% from its expected value might be caused by Neanderthal, Denisova and perhaps even "Ron's grandfather" interbreeding.

Table 2. Fit of a single "species tree" hierarchical coalescent model to the autosomal spectrum of informative characters sorted by cell-wise $X^2$ fit for the minimum $X^2$ model. The second set of values are those predicted by minimizing $G^{2"}$ starting from the more distant parameter values. The colors correspond to the particular types of pattern emphasized in figure 7.

| Pattern | Obs. | Exp($X^2$) | $X^2$ | Exp($G^{2"}$) | $X^2$ | Sign |
|---|---|---|---|---|---|---|
| NH | 1213 | 669.3 | 441.6 | 650.7 | 485.8 | 1 |
| NP | 1172 | 669.3 | 377.5 | 647.7 | 424.4 | 1 |
| NF | 1165 | 698.3 | 311.8 | 684.6 | 337.1 | 1 |
| FHP | 5340 | 4197.6 | 310.9 | 4400.9 | 200.4 | 1 |
| DP | 1071 | 669.3 | 241.0 | 652.9 | 267.8 | 1 |
| DNP | 1116 | 824.7 | 102.9 | 787.6 | 137.0 | 1 |
| SY | 3862 | 3281.5 | 102.7 | 3454.1 | 48.2 | 1 |
| DSYFHP | 4069 | 4624.7 | 66.8 | 4640.0 | 70.3 | -1 |
| SF | 2432 | 2061.4 | 66.6 | 2155.3 | 35.5 | 1 |
| YF | 3141 | 2765.9 | 50.9 | 2805.3 | 40.2 | 1 |
| DYFHP | 866 | 1093.0 | 47.1 | 1074.0 | 40.3 | -1 |
| SYFH | 1643 | 1389.5 | 46.2 | 1374.9 | 52.3 | 1 |
| YFH | 2064 | 1778.6 | 45.8 | 1827.0 | 30.7 | 1 |
| DNYFP | 413 | 563.8 | 40.3 | 547.6 | 33.1 | -1 |
| DFHP | 513 | 673.2 | 38.1 | 619.2 | 18.2 | -1 |
| DSHP | 218 | 329.5 | 37.8 | 311.0 | 27.8 | -1 |
| NSFP | 204 | 310.0 | 36.2 | 292.1 | 26.6 | -1 |
| DYFP | 245 | 355.3 | 34.2 | 324.0 | 19.3 | -1 |
| YFHP | 4234 | 3873.4 | 33.6 | 3889.0 | 30.6 | 1 |
| SYP | 1014 | 1215.0 | 33.2 | 1223.7 | 35.9 | -1 |
| DYFH | 241 | 346.3 | 32.0 | 326.5 | 22.4 | -1 |
| DNSYFH | 1179 | 1013.8 | 26.9 | 990.9 | 35.7 | 1 |
| DNSFHP | 1866 | 1657.5 | 26.2 | 1646.4 | 29.3 | 1 |
| NSHP | 237 | 329.5 | 26.0 | 309.8 | 17.1 | -1 |
| YH | 2912 | 2651.6 | 25.6 | 2682.8 | 19.6 | 1 |
| SYFP | 1204 | 1390.3 | 25.0 | 1377.5 | 21.9 | -1 |
| NYFHP | 931 | 1093.0 | 24.0 | 1077.3 | 19.9 | -1 |
| DNSFH | 402 | 512.3 | 23.7 | 528.1 | 30.1 | -1 |
| DNYFHP | 2456 | 2227.2 | 23.5 | 2254.4 | 18.0 | 1 |
| DYH | 261 | 351.0 | 23.1 | 332.1 | 15.2 | -1 |
| DNSY | 898 | 1052.4 | 22.7 | 999.6 | 10.3 | -1 |
| DN | 11849 | 12374.3 | 22.3 | 12548.4 | 39.0 | -1 |
| NYFP | 267 | 355.3 | 21.9 | 325.1 | 10.4 | -1 |
| FP | 4601 | 4928.8 | 21.8 | 4930.8 | 22.1 | -1 |
| DSF | 286 | 376.1 | 21.6 | 357.0 | 14.1 | -1 |
| DNS | 1666 | 1860.2 | 20.3 | 1867.9 | 21.8 | -1 |

Table 2 sorts the deviance on a per pattern basis. Any cell with a deviance of greater than 20 under the minimum $X^2$ criterion is shown (also shown are the worst fit cells with $G^2$, to show they are generally the same). The observed values of the patterns NH, NP and NF are all about twice as large as the model expects. This excess is consistent with the hypothesis that Neanderthals mixed with the out of Africa people. In contrast, the value for FHP is also markedly underestimated by the model.

This may be direct consequence of the model shortening the species tree internode FHP in order to better explain the high frequencies of the afore mentioned patterns. The next two most deviant patterns are those associated with the Denisovan mixing with Papuan ancestors. Note the high frequency of the DNP pattern, which may be due to the Denisovan relatives that mixed not being closely related to the Denisovan sampled.

It is interesting to consider if the evidence suggests one mixing of the our of Africa moderns with Neanderthals, or private mixing for each of the three lineages. This is unclear, since the single species tree model is distorted and it is hard to predict the exact frequencies expected. However, the shared pattern NFHP (not shown) has a deviance of 19.4 and is underrepresented by this model, while other mixing patterns such as NFH are also not showing up as highly overexpressed. While this leaves open the door to multiple mixings, the evenness of the pairwise patterns frequencies plus the difficulty of imagining how the Han and Papuans would have mixed with Neanderthals separately, is in favor of just one mixing. The lower frequency of NF compared to NH corroborates the view there is no evidence of further surviving gene infusion from Neanderthals in Europe. Thus, *Homo froggieii* is apparently no more Neanderthal than the gracile *Homo han*.

The SY pattern is the next most deviant. This may be due to introgression between these two populations, but it might also be due to sequencing errors. The pattern isolating Neanderthals from the rest of the species is underrepresented. The following two patterns of African with French are overrepresented, again perhaps reflecting low levels of gene-flow over long periods due to relative proximity, or sequencing errors. A lot of the other patterns show deficiencies and these may be due, at least in part to Papuans having 10% or more of their genome derived from archaic species of *Homo*. It is dangerous to over interpret this spectrum since the model is clearly quite tortured with a total deviance about 20 times as large as it should be due to that predicted by intra-chromosomal comparisons, such as table 1. One interesting pattern missing from table 2 is NSYFHP which supports the hypothesis of *denisova* mixing with *erectus*. This is a relatively common pattern that is only overrepresented by about 100 out of 4800, giving a small signed deviance of about +5.

**3.7 Robust modeling**

As shown in Waddell (1995), inadequacies of the model may act upon deviant patterns with high leverage to distort parameter estimates. In the case of the present data set, it is clear that there are signals strongly deviating from the single species tree coalescent model due to a mixing of Neanderthals with the modern out of Africa people, and a Denisova relative mixing Papuan ancestors. Likelihood and minimum $X^2$ fitting give high leverage to rarer patterns, such as those specifically produced by these mixing events. One way to reduce this effect is to minimize the absolute deviation divided by the expected value, or even simply minimize the sum of the absolute deviations. The former is basically minimizing the mean percent error. Doing this the best fit is 13.14% and the parameter values are 0.023445, 0.159497, 0.101626, 0.379833, and 0.333237, respectively. The values deepest in the tree have increased markedly due to the fit not being so concerned about retaining deep ancestral polymorphism to explain patterns such as NP or DP. However, minimizing just the sum of the absolute deviations, the average deviation per cell becomes 126.68, and the parameter values return to some thing like those seen with likelihood, namely 0.013212, 0.179201, 0.112174, 0.339766 and 0.243950.

Another way to gauge the effect of highly leveraged outliers, without access to a better model, is to completely deleverage the worst of the disagreeing site patterns and fit only the remaining partial spectrum. In this case, we completely deleverage the patterns NP, NH, NF, DP and DNP since these are the ones most implicated in the interspecies breeding (we do this by subtracting their deviance from the total deviance; a similar solution is obtained later if we renormalize the sum of the expected values of site patterns still being weighted to equal the observed sum of these patterns). Reestimating we obtain a fit of 1756.0 and parameter estimates of 0.032526, 0.185391, 0.100423, 0.361597 and 0.286615. If we also turn of the other patterns amongst the 10 most deviant (FHP, SY, DSYFHP, SF and YF) we end up with a fit of 1263.0 and

parameter values of 0.025549, 0.170123, 0.106303, 0.360353 and 0.277789 which remain higher than those found previously for the standard minimum $X^2$ fit. The overall fit however, remains far worse than expected by inter-chromosomal variability, so the misfit of the data may be markedly more complex than anticipated even with a two event reticulate model.

In the above examples, we did not renormalize the expected vector to sum to the exact value of the observed values being fitted. Doing this, the estimates remain close and are again increased over what they were previously with a fit of 1854.8 and parameter estimates of 0.032808, 0.186022, 0.100740, 0.362463 and 0.287183.

Combining both a robust fit criteria, absolute percent standard deviation and ignoring the NP, NH, NF, DP and DNP cells offers an insight into what else may not be fitting well. Doing this the absolute percent standard deviation drops from 13.14% to 9.85%, with parameter values of 0.037276, 0.170392, 0.102488, 0.403647, and 0.316599, which are about as large as seen in any of the species tree-fitting models consider so far. The residual remains large though compared to the 3.5% or so we are shooting for in the long run. The most deviant cells are now DNSYFH (28.1% over), NFH 27.3%, NSYFH 26.9%, DSHP -25.9%, NSFP -24.5%, DNSFHP 22.2%, NY 22.0%, FHP 22.0%, DYFP -21.9% and DNYFP -21.6%. Two of these overexpressed patterns (DNSYFH, NSYFH) might suggest that Papuan and Denisova are harboring ancient alleles. The pattern FHP is still underestimated. This may be due to the tips below (F, H and P) showing more diversity in fixation patterns than expected when FHP is fitted well (which may be an enduring issue with Papuan getting a fair few genes from Denisova, or perhaps complex mixing between the out of Africa populations). One of these patterns, NY, coincides with a hypothesis that Reich et al. (2010) were suspicious of, that is, that some Neanderthal genes may have gotten back into Africa. Supporting the possibility of this, the NS pattern is also higher than it should be (by 17%). When deviance is the measure of misfit but parameters are estimated by robust optimization, DN (signed deviance -365.7), SYFHP -287.3, FHP +211.8, SY +78.8, DNSFHP +75.3, DSYFHP -73.7, DNSYFH +72.5, DNYFHP +54.5, SF +49.4 and NY +45.0 top the list of deviants.

The direction of the patterns DN and SYFHP suggests these edge lengths may now be overestimated due to "less homoplasy" (more technically, "less polymorphism") than the model expects occurring below these edges. The patterns SY and SF (with YF just off the list) suggest post-modern mixing within Africa and Europe and / or sequencing errors. The underestimate of patterns DNSFHP, DNSYFH and DNYFHP (with DNSYFP and DNSYHP appearing not much further down the list) are all suggestive of either sequencing error or convergent mutations in Y, N, P, S, H and F respectively, in the observed data. This is because the longest pairs of edges in this tree are chimp with the terminal edges to each of these species; the patterns most readily generated by sequencing error or homoplasy. In contrast NSFYHP now has a very low deviance from the model, in agreement with the analyses in Reich et al. that suggest a very low rate of sequence error in Denisova. These are the patterns that are expected to better fit the model when the additional correction of a finite sites model is used with a hierarchical coalescent model (Waddell 1995). However, the pattern DSYFHP (generated by an apparent sequencing error in Neanderthal) is considered likely if Neanderthal has an abnormally high sequencing error rate, however, this pattern is clearly underestimated in the data.

These results with more robust fit statistics suggest that parameter values are not only being screwed up by Neanderthal and Denisovan screwing, but also, at least partly by sequencing error. There is, however, a limit to how much can be inferred from a model with a lot more misfit than is expected. Here we have made some predictions of what might be going on, in the hope that these may be tested as more complex models become available.

**3.8 Parameter free coalescent model linear invariants**

Invariants are functions that should take a particular value irrespective of the parameters used in the model. They therefor provide a useful test of some of the most fundamental aspects of a model, such as that the data are described by coalescence around a single species tree. A very useful form of invariants are linear functions of the input data. They allow the user to gauge potential fit before running the specific cases of the models they apply to. For hierarchically structured coalescences, assuming the infinite sites model, these are identified in Waddell (1995) with an exact binomial test, while a $X^2$ approximation is described in Waddell et al. (2000, 2001). For example, on the rooted three species tree 011 (that is species 1 and 2 sister to each other) and assuming the infinite sites model, and either assuming that all the external lineages have fixed all the polymorphic sites or randomly sampling to create this effect (as is the case with the pseudo-homozygotes generated by the data collection/processing procedure of Reich et al. 2010), then the frequency of site patterns 101 and 110 should be equal. That is, f(101) = f(110), so E[f(101)-f(110)] = 0 is true for hierarchically structured coalescent models. Assuming all sites are unlinked then a binomial exact test can be used to statistically test this statement with sampled data. When using the $X^2$ version of the test, the sum of the test statistics should asymptotically have a chi-square distribution with degrees of freedom equal to the sum of the degrees of freedom of the tests if they are operating on disjoint sets of patterns (when they are not operating on disjointed sets of site patterns then their expected value remains the same, but the covariance structure becomes more complicated, but still predictable from the observed site pattern frequencies). These tests allow a look at problems with the single species tree model that are insensitive to the nature of the population size fluctuations, for example, but basically say the data does not fit a single species tree. Note, if there is significant structure within an ancestral lineage, this too violates the assumption of a single species tree, according to the phylogenetic species concept, since these sub-divided populations will be fixing different alleles at different rates.

For the seven species tree ((D,N),(S,(Y,(F,(H,P))))) there are many invariants on the full site pattern spectrum induced by the two cherries of this tree. We have coded in PERL a simple algorithm to identify all the simple linear invariants induced by cherries of a tree. In table 3 the most deviant pairwise invariants are sorted and listed. Only those larger with deviance > 6 are shown, since this is roughly the critical 95% value of a chi-square with 1 degree of freedom scaled by 1.5 to account for probable linkage (likewise, later only quartets with a deviance of at least 1.5 times as large as the 95% critical value of a chi-square with 3 degrees of freedom are shown). For this tree there are 30 pairwise invariants and 8 quartet invariants. Their combined deviances are 465.4 and 313.3, respectively. Together they are about a quarter of the total deviance recorded for the fit statistic when a hierarchical coalescent model was fitted in the previous section.

The pairwise linear invariants identify the largest deviations being the pair NF versus DF. This is consistent with French getting genes from Neanderthals and not Denisovans (Reich et al. 2010). The next pattern shows a much lower frequency of the pattern of an allele shared by San, Yoruba or French with Papuans, compared to with Han. This is consistent with Papuans getting a unique infusion of archaic genes. The next pattern is not so specific, but shows that Denisovan generally shares fewer genes with the modern humans sampled than do Neanderthals (one of these patterns is consistent with the *H. denisova / H. erectus* hypothesis, that is the pattern NSYFHP). The other invariants tend to suggest either that modern human allelic patterns are rarer when Papuan is exchanged for Han, or else emphasize Denisova patterns being more common with Papuans while Neanderthal patterns are relatively more common with Han. Overall, these invariants are consistent with the hypotheses in Reich et al. (2010) of Neanderthal genes flowing into the out of Africa population and Denisova genes flowing uniquely into the Papuan lineage.

The full spectrum quartet invariants in table 3 are even more discerning per pattern than the pairwise invariants. The first shows that the DH pattern is much rarer than its equivalents involving Neanderthals and Papuans, consistent with Denisovan genes flowing only into Papuans. The deviance of the next pattern seems to be an echo of this same feature when the "bystander" states of 1 in San and French are added in, except that now it is an excess of Neanderthal with moderns except Papuans. This is an interesting subtlety. If Neanderthals had spread their genes evenly amongst the out of Africa people, this would not be expected. There may be some subtle evidence of French and Han having some Neanderthal genes that the Papuans do not have. The Papuan might have lost up to 8% or so of its genes originating on the out of Africa Neanderthal mixing by being overwritten by the later Denisovan mixing, but the differences observed here seem larger than that in percentage terms.

Table 3. Simple linear invariants for coalescent models with the species tree ((D,N),(S,(Y,(F,(H,P))))) with the infinite sites assumption. The first three columns track the pairwise only invariants, while the columns on the right are quartet linear invariants. These are assessed it the sixth column based on arbitrary pairs determined by the numerical order of the patterns (using binary indexing). The seventh column gives the deviance of each pattern within the quartet, while the final column is the deviance of the quartet overall.

| Pattern | Obs | $X_1^2$ | Pattern | obs | $X_1^2$ | $X_1^2$ | $X_3^2$ |
|---|---|---|---|---|---|---|---|
| NF | 1165 | 171.4 | NH | 1213 | 177.9 | 35.0 | 203.2 |
| DF | 613 | | DH | 639 | | 144.6 | |
| SYFP | 1204 | 67.7 | NP | 1172 | 4.5 | 21.5 | |
| SYFH | 1643 | | DP | 1071 | | 2.2 | |
| NSYFHP | 4776 | 56.5 | NSFH | 325 | 13.8 | 22.3 | 32.5 |
| DSYFHP | 4069 | | DSFH | 237 | | 0.7 | |
| DNP | 1116 | 35.1 | NSFP | 204 | 2.2 | 8.5 | |
| DNH | 853 | | DSFP | 235 | | 0.9 | |
| FP | 4601 | 28.4 | NSYFH | 502 | 6.7 | 15.7 | 25.9 |
| FH | 5127 | | DSYFH | 423 | | 0.0 | |
| YP | 2523 | 27.8 | NSYFP | 359 | 2.1 | 9.1 | |
| YH | 2912 | | DSYFP | 399 | | 1.1 | |
| YFP | 1777 | 21.4 | NFH | 429 | 15.6 | 13.9 | 19.4 |
| YFH | 2064 | | DFH | 321 | | 3.9 | |
| SFP | 1210 | 19.3 | NFP | 347 | 0.1 | 0.4 | |
| SFH | 1436 | | DFP | 337 | | 1.3 | |
| SP | 1966 | 7.0 | NYFH | 320 | 10.0 | 10.0 | 14.8 |
| SH | 2136 | | DYFP | 245 | | 2.0 | |
| SYP | 1014 | 7.0 | NYFP | 267 | 1.3 | 0.0 | |
| SYH | 1137 | | DYFH | 241 | | 2.8 | |
| NSF | 350 | 6.4 | | | | | |
| DSF | 286 | | | | | | |

Another set of invariants, the three informative site patterns induced by rooted triples, are described in detail in Waddell et al. (2000 and 2001). Table 4 shows the result of applying the full set of these invariants to the current data set. Also shown is the $P_1$ statistic of Waddell et al. (2000, 2001) that measures the likelihood ratio of the rooted binary triple suggested by the species tree, verses the null hypothesis of a rooted trinary triple (an unresolved three taxon rooted tree). Note that the worst fit by the $P_2$ statistic involves the triple ((D,N),H). The $P_1$ statistic is very large and significant and indicates that this is a highly resolved triple consistent with the species tree of figure 1a, but the two patterns DH and NH are significantly different in frequency, when they should be equal under a hierarchically structured coalescent model based on a single species tree. Ditto for the second most deviant pattern. The third strongest violator of the expected equality of the non-species tree informative sub-patterns is the triple (F,(H,P)). This triple is much less strongly resolved and there is a substantial deficiency of the FP patterns in the data. This triple is consistent with suggestions that Han may be a approximately 50:50 hybrid of members of the French and Papuan lineages, although later we will look at this more closely and consider other possibilities. The misfit of triples such as (Y,(H,P)) to (S,(F,P)) may be part of the proximal cause of the NeighborNet splitting SYFH from all the others.

It is interesting to note that the rooted triples for ((D,N),Y) and ((D,N),S) are both of a similar degree of misfit on $P_2$ in favor of N with modern. This might indicate a member of the Neanderthal lineage mixing with the ancestor of all moderns, but it might also be Denisova picking up a significant fraction of genes that lie outside the Neanderthal, Denisova, modern human, clade. This has not been confirmed yet, but is consistent with the rather old mtDNA form found in Denisova (Krause et al. 2010). Here, there is little evidence here for NY being better supported than NS, a possibility considered by Reich et al. (2010).

Finally, it is interesting how little evidence there is for specific mixing between, for example, San, Yoruba and Han. It would be very much anticipated, *a priori* that the San and Yoruba, being relatively close to each other for a very long period of time, may have experienced appreciable gene flow. In contrast, it is very hard to see how San and Han would experience appreciable gene flow since the out of Africa event, perhaps 60,000 years ago. However, the data seem to suggest that the SY signal is actually weaker than the SH signal. Thus, the general misfit of the model does not seem to point towards gene flow within Africa. It will be important to work with a much-improved version of this data set (e.g. confidently calling diploid versus haploid states) to be confident errors are not hiding these finer details from view.

Table 4. The full set of rooted triples for the species tree in figure 1a under an infinite sites hierarchical coalescent model for all sub spectra of rooted 3 species trees. The $P_1$ statistic of Waddell, Kishino and Ota (2001) is the $G^2$ likelihood ratio of this data under the species tree hypothesis versus a three-way coalescent. The $P_2$ statistic of Waddell, Kishino and Ota (2001) is the $X^2$ statistic for the deviation from the expected equality of the two nonspecies tree signals. The results are sorted by the magnitude of the P2 statistic's deviation from expectation. Here, x is the leftmost of the two most closely related lineages, y is its sister lineage, and z is their mutual sister.

| Rooted Triple | $f_{xy}$ | $f_{xz}$ | $f_{yz}$ | $P_1$ statistic | $P_2$ statistic |
|---|---|---|---|---|---|
| ((D,N),H): | DN(23176), | DH(10064), | NH(11980), | P1= 3069.57, | P2= 166.53 |
| ((D,N),F): | DN(23323), | DF(10026), | NF(11851), | P1= 3186.18, | P2= 152.24 |
| (F,(H,P)): | HP(15601), | FH(15580), | FP(13604), | P1=   22.57, | P2= 133.79 |
| (Y,(H,P)): | HP(22385), | YH(12963), | YP(11427), | P1= 2102.76, | P2=  96.73 |
| (Y,(F,P)): | FP(21015), | YF(13557), | YP(12054), | P1= 1379.91, | P2=  88.20 |
| (S,(H,P)): | HP(25784), | SH(11624), | SP(10277), | P1= 4312.26, | P2=  82.85 |
| (S,(F,P)): | FP(24359), | SF(12173), | SP(10849), | P1= 3272.57, | P2=  76.14 |
| (N,(S,H)): | SH(26192), | NS( 9949), | NH(11199), | P1= 4805.27, | P2=  73.88 |
| (N,(Y,H)): | YH(30064), | NY( 9243), | NH(10333), | P1= 7674.59, | P2=  60.69 |
| ((D,N),Y): | DN(23580), | DY(10290), | NY(11294), | P1= 3395.66, | P2=  46.70 |
| (N,(S,F)): | SF(26271), | NS(10051), | NF(11025), | P1= 4874.36, | P2=  45.01 |
| ((D,N),S): | DN(23670), | DS(10256), | NS(11224), | P1= 3471.08, | P2=  43.62 |
| (D,(S,P)): | SP(25885), | DS(10021), | DP(10969), | P1= 4713.07, | P2=  42.82 |
| (D,(F,P)): | FP(36955), | DF( 7698), | DP( 8529), | P1=14464.22, | P2=  42.56 |
| ((D,N),P): | DN(23028), | DP(10562), | NP(11497), | P1= 3001.49, | P2=  39.63 |
| (N,(S,P)): | SP(25240), | NS(10344), | NP(11259), | P1= 4160.48, | P2=  38.76 |
| (S,(Y,P)): | YP(17866), | SY(14333), | SP(13317), | P1=  349.45, | P2=  37.33 |
| (D,(Y,P)): | YP(29598), | DY( 9309), | DP(10133), | P1= 7467.13, | P2=  34.92 |
| (N,(Y,F)): | YF(30087), | NY( 9299), | NF(10113), | P1= 7774.93, | P2=  34.13 |
| (N,(Y,P)): | YP(28861), | NY( 9576), | NP(10331), | P1= 6801.92, | P2=  28.63 |
| (D,(H,P)): | HP(38596), | DH( 7527), | DP( 8173), | P1=16114.55, | P2=  26.58 |
| (N,(H,P)): | HP(37701), | NH( 8548), | NP( 8213), | P1=14595.75, | P2=   6.70 |
| (N,(F,H)): | FH(37626), | NF( 8218), | NH( 8494), | P1=14578.83, | P2=   4.56 |
| (D,(S,H)): | SH(27147), | DS( 9936), | DH(10238), | P1= 5719.09, | P2=   4.52 |
| (S,(Y,H)): | YH(18596), | SY(13527), | SH(13858), | P1=  507.03, | P2=   4.00 |
| (S,(Y,F)): | YF(18525), | SY(13489), | SF(13797), | P1=  504.47, | P2=   3.48 |
| (D,(F,H)): | FH(38832), | DF( 7599), | DH( 7784), | P1=16561.96, | P2=   2.22 |
| (D,(Y,H)): | YH(31032), | DY( 9207), | DH( 9385), | P1= 8809.83, | P2= 1.70 |
| (N,(S,Y)): | SY(26096), | NS(10184), | NY(10344), | P1= 5008.75, | P2= 1.25 |
| (D,(S,Y)): | SY(26790), | DS( 9910), | DY(10034), | P1= 5628.12, | P2= 0.77 |
| (D,(S,F)): | SF(27188), | DS(10000), | DF(10117), | P1= 5766.95, | P2= 0.68 |
| (N,(F,P)): | FP(36142), | NF( 8710), | NP( 8651), | P1=13027.46, | P2= 0.20 |
| (Y,(F,H)): | FH(21915), | YF(12481), | YH(12514), | P1= 1796.34, | P2= 0.04 |
| (S,(F,H)): | FH(25322), | SF(11160), | SH(11183), | P1= 3932.66, | P2= 0.02 |
| (D,(Y,F)): | YF(31041), | DY( 9249), | DF( 9242), | P1= 8873.88, | P2= 0.00 |

The rooted triple tests of Waddell et al. (2000, 2001) are readily extended to rooted quartets. The off species tree sub-spectral invariants of these quartets are shown in table 5. They give a bit more discrimination that the triples. The first five most significant results point out the generality of the pattern of Papuan sharing more alleles with Denisovan and markedly less derived alleles than expected with other modern humans. The next set of derived alleles on symmetric rooted quartets suggests the pattern that Papuan shares exclusive derived alleles with Denisova is too high, Neanderthal sharing alleles with Han or French is too high, while Denisovan with French or Han is too low, yet Neanderthal with Papuan is relatively, neither too high nor too low.

Table 5. The full set of rooted quartets for the species tree of figure 1a under an infinite sites hierarchical coalescent model of all sub spectra for rooted 4 species trees. The frequency (f) of each invariant pattern is shown, along with its signed $P_2$ $X^2$ test statistic, followed by the total $P_2$ statistic for that quartet and finally, the degrees of freedom to be used when formally testing against the null hypothesis. To clarify, in the first row, $f_{13}$ is the allelic pattern 1010, where 0 is the ancestral state, thus D and F alone show the derived state. Similarly, for the first row, $f_{14}$ is pattern DP, $f_{23}$ is YF, while $f_{24}$ is YP. The signed $P_2$ statistic of Waddell, Kishino and Ota (2001) is the $X^2$ statistic for the deviation from the expected value of these invariants of the hierarchically structured single species tree coalescent model. The results are sorted by the magnitude of the total $P_2$ statistic's deviation from expectation.

| Rooted Quartet | $f_{13}$ | ($P_2$) | $f_{14}$ | ($P_2$) | $f_{23}$ | ($P_2$) | $f_{24}$ | ($P_2$) | $P_2$total | df |
|---|---|---|---|---|---|---|---|---|---|---|
| (D(Y(FP))) | 3743 | -47.4 | 4634 | 47.4 | 9602 | 58.6 | 8159 | -58.6 | 212.01 | 2 |
| (D(S(FP))) | 3814 | -42.8 | 4666 | 42.8 | 8289 | 55.6 | 6986 | -55.6 | 196.75 | 2 |
| (D(Y(HP))) | 3698 | -34.4 | 4446 | 34.4 | 9134 | 61.1 | 7700 | -61.1 | 190.86 | 2 |
| (D(S(HP))) | 3658 | -33.2 | 4389 | 33.2 | 7755 | 55.9 | 6493 | -55.9 | 178.19 | 2 |
| ((DN)(HP)) | 2661 | -83.3 | 3159 | -0.1 | 3682 | 80.9 | 3199 | 0.2 | 164.42 | 3 |
| ((DN)(FP)) | 2712 | -93.2 | 3248 | -0.1 | 3724 | 65.0 | 3370 | 3.5 | 161.73 | 3 |
| ((DN)(FH)) | 2706 | -40.4 | 2744 | -32.1 | 3325 | 23.5 | 3454 | 51.5 | 147.39 | 3 |
| ((DN)(SH)) | 3722 | -5.6 | 3530 | -29.8 | 3735 | -4.7 | 4491 | 99.8 | 139.91 | 3 |
| ((DN)(YH)) | 3475 | -4.9 | 3249 | -35.7 | 3511 | -2.6 | 4197 | 96.2 | 139.38 | 3 |
| (D(S(YP))) | 4525 | -43.8 | 5460 | 43.8 | 9549 | 22.5 | 8644 | -22.5 | 132.57 | 2 |
| (N(S(YP))) | 4720 | -40.4 | 5635 | 40.4 | 9477 | 20.2 | 8621 | -20.2 | 121.34 | 2 |
| ((DN)(SF)) | 3671 | -4.1 | 3441 | -33.2 | 3722 | -1.4 | 4349 | 80.6 | 119.33 | 3 |
| ((DN)(YF)) | 3456 | -2.7 | 3192 | -36.9 | 3506 | -0.7 | 4063 | 72.8 | 113.11 | 3 |
| (N(F(HP))) | 4460 | -1.4 | 4617 | 1.4 | 11492 | 51.2 | 10008 | -51.2 | 105.15 | 2 |
| (N(Y(FP))) | 4555 | -2.6 | 4773 | 2.6 | 9402 | 42.8 | 8176 | -42.8 | 90.6 | 2 |
| (N(Y(HP))) | 4529 | 0.0 | 4527 | 0.0 | 8944 | 43.4 | 7741 | -43.4 | 86.74 | 2 |
| (N(S(FP))) | 4579 | -2.9 | 4813 | 2.9 | 8042 | 35.3 | 7011 | -35.3 | 76.44 | 2 |
| (N(S(YH))) | 4622 | -36.2 | 5477 | 36.2 | 8906 | -0.3 | 9002 | 0.3 | 72.9 | 2 |
| (S(Y(HP))) | 5654 | 13.6 | 5113 | -13.6 | 6993 | 20.1 | 6263 | -20.1 | 67.38 | 2 |
| (Y(F(HP))) | 6125 | 9.0 | 5665 | -9.0 | 8742 | 24.4 | 7842 | -24.4 | 66.79 | 2 |

| | | | | | | | | | |
|---|---|---|---|---|---|---|---|---|---|
| (N(S(HP))) | 4475 | -0.2 | 4535 | 0.2 | 7551 | 32.0 | 6599 | -32.0 | 64.45 | 2 |
| (S(F(HP))) | 5477 | 5.3 | 5143 | -5.3 | 9433 | 25.9 | 8470 | -25.9 | 62.3 | 2 |
| (S(Y(FP))) | 5952 | 10.1 | 5472 | -10.1 | 7336 | 15.5 | 6677 | -15.5 | 51.16 | 2 |
| (N(S(YF))) | 4692 | -23.0 | 5373 | 23.0 | 8882 | -0.9 | 9057 | 0.9 | 47.78 | 2 |
| ((DN)(SP)) | 3716 | -23.3 | 4022 | 0.0 | 4039 | 0.1 | 4312 | 20.9 | 44.26 | 3 |
| ((DN)(YP)) | 3440 | -17.2 | 3712 | 0.1 | 3707 | 0.1 | 3910 | 12.8 | 30.24 | 3 |
| ((DN)(SY)) | 3576 | -7.1 | 3610 | -4.5 | 3850 | 3.3 | 3920 | 8.8 | 23.61 | 3 |
| (D(S(YH))) | 4527 | -2.3 | 4731 | 2.3 | 8847 | -3.5 | 9204 | 3.5 | 11.56 | 2 |
| (D(S(YF))) | 4574 | -0.4 | 4657 | 0.4 | 8814 | -4.4 | 9212 | 4.4 | 9.53 | 2 |
| (N(Y(FH))) | 4393 | -2.7 | 4613 | 2.7 | 8656 | 0.0 | 8633 | 0.0 | 5.4 | 2 |
| (N(S(FH))) | 4409 | -1.7 | 4583 | 1.7 | 7351 | 0.2 | 7272 | -0.2 | 3.79 | 2 |
| (D(Y(FH))) | 3798 | -1.3 | 3941 | 1.3 | 8680 | 0.0 | 8671 | 0.0 | 2.65 | 2 |
| (D(S(FH))) | 3799 | -1.0 | 3920 | 1.0 | 7360 | 0.1 | 7319 | -0.1 | 2.01 | 2 |
| (S(Y(FH))) | 5626 | -0.2 | 5687 | 0.2 | 6947 | -0.2 | 7018 | 0.2 | 0.69 | 2 |

### 3.9 An infinite sites reticulate model

The first reticulate model considered is that of Neanderthal mixing only once with the out of Africa group. This model is based on the models described in Waddell (1995), where hybridization can be described as a sum of the spectra of distinct binary trees. In this case, the spectrum of each "tree" is in fact produced by a hierarchically structured coalescent model following a single species tree. This is described further in figure 9.

Fitting this model produces a minimum $X^2$ fit of 2554.51 with parameter values of g1 = 0.029130, g2 = 0.167078, g3 = 0.116756, g4 = 0.371688, g5 = 0.283454, g6 = 0.25153, g8 = 0.0 and P(N) = 0.054453. Turning off the signals DP and DNP that are most strongly associated with a Denisovan mixing with Papuan, then the minimum $X^2$ fit of 2178.71 is achieved with parameter values of g1 = 0.033921 (0.027455-0.040440), g2 = 0.169044 (0.163607-0.174520), g3 = 0.116076 (0.110668-0.121526), g4 = 0.375043 (0.368325-0.381804), g5 = 0.282431 (0.274312-0.290606), g6 = 0.383276 (0.317097-0.452853), g8 = 0 (0-0.008409), and P(N) = 0.046551 (0.042092-0.051083). The numbers in the brackets are a 95% confidence interval based on the point where the fit becomes 6 units worse than the optimum, when all other parameter values are held constant (which is basically the 95% point of a chi-square with 1 d.f. multiplied by the inflation factor 1.5 due to linkage). Thus, the fit of this model has improved considerably over that of a single species tree model.

The parameter estimates for g1-g5 of the single Neanderthal introgression reticulate model are higher than those of the single species model except for g2, but they are all fairly similar to previous estimates. The new parameter g6 takes a moderate size in both versions of the model. This in turn suggests that the effective population size leading from the common ancestor of modern humans and Neanderthals down to the Neanderthals of the last 100,000 years in both Europe and the middle East (which is where the Neanderthals that mixed with all the out of Africa people would seem to need to be located) was only slightly smaller than that leading to modern humans heading out of Africa. This may be in disagreement with Reich et al. who suggest a very strong constriction in the Neanderthal lineage. However, they are comparing European Neanderthals of the last 50 thousand years or so Europe, and not a Middle Eastern population of around 50 to 80 thousand years ago. The proportion of genes coming from Neanderthals into modern humans is around 5%, which agrees well with estimates in Reich et al. (2010). The data/model is unable to reject the hypothesis that the Neanderthal population which breed with modern humans was a direct ancestor of the European Neanderthals sequenced (since

g8 takes a value close to zero).

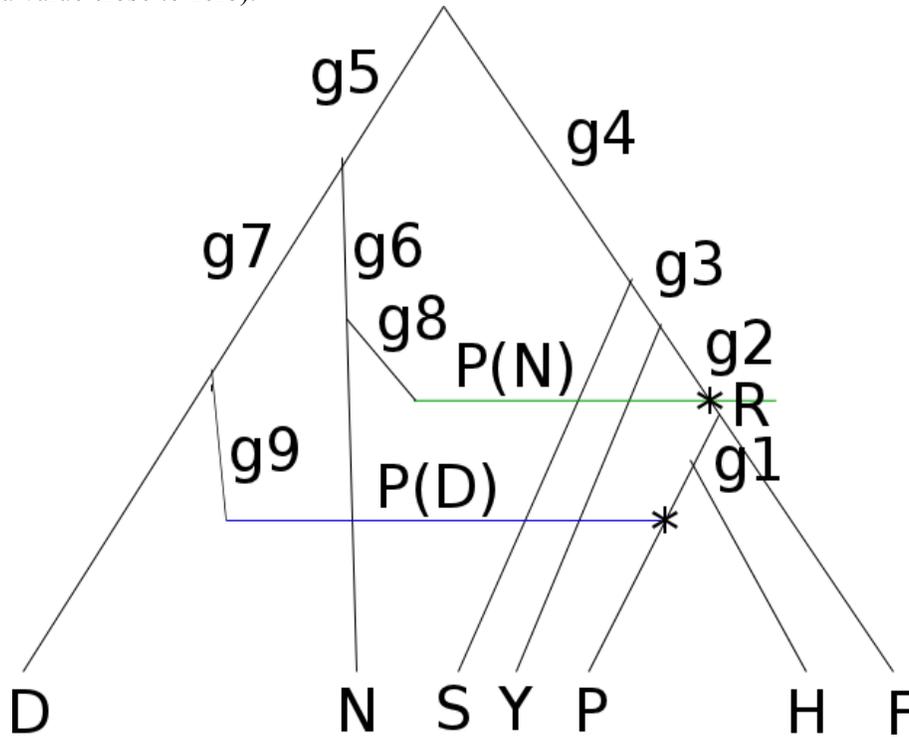

Figure 9. The reticulate model we use. Parameters g1-g9 are measured in units of the number of generations divided by the effective population size (males and females together). The asterisks mark the reticulate interbreeding points, P(D) is the proportion of genes contributed by the Denisovan relative to the Papuan population, while P(N) is the proportion of genes for the out of Africa people contributed by a Neanderthal population. R is a parameter that only appears if it is known the Neanderthal population contributing to the out of Africa people is directly ancestral to the sequenced Neanderthal, else parameter g8 is necessary. All terminal edges are effectively infinite in length as full coalescence in that period is forced upon them by the process of scoring the sequence variants to create pseudo-homozygotes. Following from Waddell (1995) this model induces four species trees (which could also be called four hierarchical coalescent models). These four frequency spectra are induced, with probability (1-P(D))(1-P(N)) by T(s) = ((D,N):g5,(S,(Y,(F,(H,P):g1):g2):g3):g4)), with probability P(D)(1-P(N)) by T(D) = (((D,P):g7),N):g5,(S,(Y,(F,H):g2):g3):g4)), with probability P(1-P(D))P(N) by T(N) = ((D,(N,(F,(H,P):g1):g8):g6):g5,(S,Y):g4)), and finally, with probability P(D)P(N)) by T(DN) = (((D,P):g7,(N,(F,H):g8):g6):g5,(S,Y):g4)). All terminal edges are effectively infinite in length since the diploid allelic data was called as a pseudo-homozygote. Thus the edge to P in T(D) and T(DN) equals g9 plus infinity, which remains infinity, thus parameter g9 is not identifiable with this data. Each of these four trees induces a spectrum under a hierarchical coalescent model, and the reticulate informative spectrum is the sum of each f spectra (sum of all elements equals 1) multiplied by the probability of that species tree element (e.g. P(D)P(N)) for T(DN)). In general $\mathbf{s}(\text{reticulate model}) = \sum_i \mathbf{s}(T_i)$. However, when conditioning on, for example, just the informative sites, it is important to preserve information on their scale $\phi_i$ relative to all events (i.e., no change or a singleton event, in this conditioning). In that case we have $\mathbf{s}(\text{reticulate model: conditional}) = \varphi \sum_i \phi_i \mathbf{s}(T_i)$, where $\varphi$ is just a renormalization term so the complete sum equals one.

Correcting the minimum $X^2$ with the MS results at the above points, a new solution is found with $X^{2'} = 2224.74$, and parameters g1 = 0.028335, g2 = 0.167548, g3 = 0.116392, g4 =

0.371222, g5 = 0.284098, g6 = 0.246541, g8 = 0.000000, and P(N) = 0.056085. These remain close to their previous values. Correcting for the minimum $X^2$ with the DP and DPN signals off results in $X^{2'}$ = 1793.65, and parameters g1 = 0.033178, g2 = 0.169677, g3 = 0.115752, g4 = 0.374702, g5 = 0.282093, g6 = 0.377347, g8 = 0.000000, and P(N) = 0.047389. Again, these remain close to their previous values and all are comfortably within their previous 95% CI.

Next, the data is reduced to six species by removing the Papuan sequence. The optimal fit of 1276.65 is achieved with parameter values of g2 = 0.181111 (0.174079-0.188194), g3 = 0.116802 (0.110947-0.122702), g4 = 0.377066 (0.370219-0.383955), g5 = 0.288641 (0.280594-0.296741), g6 = 0.366361 (0.282191-0.454756), g8 = 0 (0-0.016258), and P(N) = 0.037692 (0.033171 - 0.042701). The numbers in the brackets are a 95% CI based on the point where the fit becomes 6 worse than the optimum (which is basically the 95% point of a chi-square with 1 d.f. multiplied by the inflation factor 1.5 due to linkage). A formal test of the hypothesis that this edge is zero may be made using a 50:50 $X_0^2 + X_1^2$ distribution due to the edge length of the tree having a boundary at zero (Ota et al. 2000) and this approach should also be used to construct a 95% CI which would run from 0 to 0.011076. The parameters remain fairly close to their expected values with the most notable change being a decrease in the proportion of Neanderthal in the data. This may be due to this parameter no longer having to explain extra DN patterns due to the hypothesized Denisova-Papuan interbreeding.

The overall fit of this model has improved considerably from the single species tree hierarchical coalescent model. However, it would be expected to have a value of $2^6$-6(singletons)-1(constants) - 1(constraint on sum of all observations) – 7 fitted parameters or 49 assuming independence. Since the actual value is over 1200, the overall model is not a close match to expectations.

The largest deviants for this model are shown in table 6. While fitting of a Neanderthal mixing event has helped the fit, there is still a poor fit in terms of the exact details. There are too few shared NFH patterns and far too many NF and NH patterns. This could be due to multiple private interbreeding events with Neanderthals. However, it might also be due to selection favoring the retention of different Neanderthal alleles in different populations or data errors. Whatever the cause, this is probably a primary cause of the edge leading from Neanderthals to the common ancestor of French and Han having zero length (indeed, it very much wants to go negative). If so, this edge being zero length may not simply be the Neanderthals of the middle East being indistinguishable from an ancestor of those in Europe.

Notice that there are further pairwise site patterns (such as SF and SY) where the observed value is greater than the expected (often by 100 to 200). This could be due to interbreeding amongst groups of modern humans. However, it does not fit geographic expectations well. For example, the pattern YF is over by nearly the same amount as YH, which has exactly the same expected value (the pair are an invariant of this model), yet biogeographically, the former would be expected to be boosted by migration far more readily than the latter. Misfits of this sort of size do not seem to be readily explained by uniform sequencing error however. Data in the supplements of Reich et al. (2010) suggest that the error rate after their filtering of the data is less than 1/1000 sites. Since long terminal edges dominate the data, the most likely way to generate the pattern YH would be two independent errors at an otherwise constant site. If we look at the raw data then, assuming independence, the probability of a double hit of YH is equal to 97234 x 89751 / 335821175 or about 26. However, if there are hotspots of either substitution or sequencing error, the frequency of double hits could climb appreciably. However, patterns such as DSYFH might seem to be even more likely to be generated by sequencing errors in this reduced data and they show an underrepresentation. Thus to clarify whether the problems in fit are likely due to sequencing errors, it is necessary to fit finite sites models explicitly. There is still some excess for the pattern placing Denisova with Chimp, but the size of its deviance is swamped by many other worse fitting patterns.

Table 6. The most deviant patterns when a single Neanderthal-mixing event is allowed fitted to the data by minimum $X^2$ with Papuan removed.

| Pattern | Observed | Expt | Deviance | Pattern | Observed | Expt | Deviance |
|---|---|---|---|---|---|---|---|
| NH | 1626 | 1200.9 | 150.4 | NY | 1445 | 1276.2 | 22.3 |
| FH | 10467 | 9501.6 | 98.1 | SFH | 4007 | 3719.9 | 22.2 |
| NFH | 988 | 1320.6 | -83.8 | DYF | 550 | 660.5 | -18.5 |
| NF | 1512 | 1200.9 | 80.6 | DNYF | 881 | 1015.1 | -17.7 |
| SF | 3642 | 3167.7 | 71.0 | DNS | 2152 | 2356.2 | -17.7 |
| DNFH | 1455 | 1791.8 | -63.3 | DYH | 555 | 660.5 | -16.8 |
| DYFH | 1107 | 1391.3 | -58.1 | NSYH | 628 | 735.6 | -15.7 |
| SH | 3614 | 3201.2 | 53.2 | DSF | 521 | 618.5 | -15.4 |
| DSYFH | 4492 | 4974.0 | -46.7 | YFH | 6298 | 6026.1 | 12.3 |
| DFH | 834 | 1045.1 | -42.6 | NSY | 807 | 910.7 | -11.8 |
| DNSYH | 1704 | 1493.1 | 29.8 | DNSFH | 2268 | 2111.9 | 11.5 |
| YF | 4918 | 4553.4 | 29.2 | DNSF | 909 | 1008.9 | -9.9 |
| YH | 4914 | 4553.4 | 28.5 | NSYFH | 5278 | 5065.8 | 8.9 |
| DN | 12965 | 13579.4 | -27.8 | DSFH | 887 | 979.4 | -8.7 |
| SYFH | 9534 | 10036.2 | -25.1 | DNYH | 923 | 1015.1 | -8.4 |
| DSH | 494 | 618.5 | -25.1 | DSYH | 661 | 735.7 | -7.6 |
| DNSYF | 1685 | 1493.1 | 24.7 | NSF | 554 | 622.2 | -7.5 |
| NYFH | 1251 | 1433.6 | -23.3 | DF | 950 | 1034.8 | -6.9 |
| DSY | 842 | 991.8 | -22.6 | NYF | 598 | 664.1 | -6.6 |

### 3.10 The Full Monty infinite sites model

This brings us to some exploration of the most complete model as yet, otherwise known as the "Full Monty", possibly derived from a Sheffield colloquialism for a professionally tailored full suite. This model allows for both Neanderthal and Denisova gene intromission as described in figure 9.

To explore this model more directly, we wrote a PERL script to control MS and to process the results into the informative site pattern spectrum. It was determined that 10 million replicates per "species" tree with MS were sufficient for our purpose. After summing up the edge lengths of these 10 million trees, the length of all the informative edge weights was also approximately 16 million. The fluctuation in the $X^2$ statistic between independent replicates of this size, rescaled to sum to the observed number of informative patterns (149,019), was about 1.22, which is less than 1/100 of the expected value for a multinomial of length 149,019 (and close to that expected for a rescaled multinomial of length 16 million). The fluctuation of the $X^2$ statistic with respect to the real data is larger, due to the poor fit of data to model (with a mean $X^2$ of closer to 1800 rather than ~120), thus the distance of simulated datasets of 10 million to the observed data had an intrinsic fluctuation of ~6. This effect of the quadratic term increasing fluctuations between replicates the further the observed is from the expected can be negated by using the robust weighted $L_1$ metric $\sum \frac{|\text{obs} - \text{exp}|}{\text{exp}}$, which is close to the ML estimator when the error on each cell is symmetrically exponentially distributed (either up or down), with s.d. equal to the size of the cell value. Starting at the parameter values found for the Neanderthal mixing model described above, we took steps of ~4.s.d. in all possible pairs of directions. Fitting a quadratic model we then stepped towards an optimum. The best $X^2$ fit we achieved had a value of ~1831, which we now diagnose. At these parameter values, either weighting and summing site pattern frequencies across species trees first then normalizing, or normalizing then weighting and summing, made very little difference (e.g., $X^2 =$ 1831.5 versus 1831.2, with very similar parameter values, respectively), most likely due to the very similar sum of informative patterns across all four species trees. Here will focus on the technically more correct approach of summing first, then normalizing.

The parameters, the $X^2$ fit and most deviant patterns at the best optimum found for the full model shown in figure 9 are given in table 7. The edge lengths of the model remain similar to those seen earlier with either a single species tree or a single Neanderthal mixing. The new parameters g6, g7 and g8 take on values of 0.28, 0.0625 and 0, respectively. It is worth noting that all these parameters have a standard error approximately 10 times as big as those on g1-g5, since they only appear on the secondary species trees, which have much lower weight than the main species tree. Provisionally, then, the value for g6 suggests a similar amount of coalescence in the Neanderthal lineage prior to mixing with the out of Africa group of moderns, as they had experience themselves from the common ancestor. Again, g8 goes to zero suggesting that we cannot distinguish the Neanderthals that mixed with moderns from the ancestors of the European Neanderthals sequenced, given this data and these models. Finally, g7 is relatively small, suggesting that the ability to clearly discriminate the creature that mixed with the ancestors of the Papuan as closer to the Denisovan lineage, rather than to the Neanderthal lineage is relatively weak. This seems quite feasible, as this creature seems to have existed in extreme Southeast Asia, a long way from Denisova in Siberia. It may also have incorporated alleles from even earlier branching *Homo* lineages independently of the Denisova lineage.

Table 7. The most deviant patterns with a single Neanderthal-mixing event and a single Denisovan-mixing event, as shown in figure 9. The total $X^2$ fit was 1831.5 with summation of internal edge vectors first, and parameter values g1:0.0415, g2:0.1765, g3:0.116, g4:0.3725, g5:0.2895, g6:0.28, g7:0.0625, g8:0, P(N):0.0446 and P(D):0.0549. Produced by simulations of 10 million gene trees for each of the four species tree components.

| Pattern | Observed | Expt | Deviance | Pattern | Observed | Expt | Deviance |
|---|---|---|---|---|---|---|---|
| NH | 1213 | 740.5 | 301.4 | YFHP | 4234 | 3963.9 | 18.4 |
| FHP | 5340 | 4354.4 | 223.1 | DFHP | 513 | 619.3 | -18.3 |
| NF | 1165 | 784.5 | 184.5 | DNH | 853 | 737.8 | 18.0 |
| DNFHP | 990 | 1281.3 | -66.2 | DNSYHP | 1120 | 991.6 | 16.6 |
| NFHP | 559 | 785.2 | -65.2 | SYP | 1014 | 1150.9 | -16.3 |
| SF | 2432 | 2101.6 | 51.9 | NSFP | 204 | 270.0 | -16.1 |
| DSYFHP | 4069 | 4494.4 | -40.3 | NHP | 413 | 500.8 | -15.4 |
| NY | 1131 | 955.4 | 32.3 | YH | 2912 | 2709.3 | 15.2 |
| DNSYFH | 1179 | 1014.9 | 26.5 | DNSFH | 402 | 487.5 | -15.0 |
| DNSYFP | 1098 | 941.5 | 26.0 | NYFHP | 931 | 1054.8 | -14.5 |
| DYFHP | 866 | 1028.5 | -25.7 | DFH | 321 | 396.3 | -14.3 |
| DNSFHP | 1866 | 1661.0 | 25.3 | SHP | 1478 | 1341.0 | 14.0 |
| DNF | 935 | 794.3 | 24.9 | SH | 2136 | 1972.5 | 13.6 |
| NFP | 347 | 451.8 | -24.3 | DSYFH | 423 | 505.1 | -13.3 |
| YF | 3141 | 2882.5 | 23.2 | NSYH | 239 | 300.5 | -12.6 |
| NP | 1172 | 1018.5 | 23.1 | DNSFP | 425 | 504.7 | -12.6 |
| DNYFP | 413 | 520.4 | -22.2 | DNFP | 483 | 564.6 | -11.8 |
| DN | 11849 | 12369.0 | -21.9 | DSY | 564 | 650.8 | -11.6 |
| SFHP | 2571 | 2356.2 | 19.6 | NSYP | 242 | 300.5 | -11.4 |

The main cause of misfit of the model seems to be effectively that identified earlier with the single Neanderthal-mixing event. There are far too many patterns of a Neanderthal allele being shared with just one of the out of Africa individuals, and generally far too few shared between archaics and more than one of the out of Africa people (including the pattern NFHP). Contradicting this, there are too many patterns shared by the out of Africa people. Again, this could be data errors creating Neanderthal and modern patterns. While the pattern NY is also overexpressed, the pattern NS is only about 45 out of 1419 too large and fits the model very well. This would seem to argue against this being solely sequencing error unless the San genome has a markedly lower rate of sequencing error than the other moderns (there is some evidence in Reich et al. 2010, to suggest this).

It is tempting to put the burden for some of this misfit shown in table 7 on the Han individual as being a potential mixture of a the lineages leading to European and Papuan, separately (e.g. Reich et al. 2011). However, patterns that would suggest this, which are FH (misfit +0.1), HP (-6.3), NFH (-3.7), NHP (-15.4) relative to patterns FP (-0.8), or NFP (-24.3), suggest that it is only the Neanderthal with pairs of out of Africa people that fit poorly. It seems tempting then to think that a model of three independent out of Africa lineages, with three independent mixings with the same population of Neanderthals (plus the independent Denisovan mixing event), would fit markedly better than the present model. The number of extra parameters would be the gain of a g01 and P(NP), g02 and P(NH) and g03 and P(NF) (while parameters g8, P(N) and g1 might all be lost). It would seem that the P parameters would likely remain in the range of 0 to 0.5, while the g0 parameters might need to be markedly non-zero to explain the extent of alleles shared privately with Neanderthals. The anticipated improvement in fit could be to a value of 1000 or better (obtained by summing up the misfit on the patterns most readily explained by this more complicated demographic model). That would seem a marked improvement for a model adding only 4 parameters or so.

What is also interesting is the very minor support, if any, this data offers the hypothesis of Han being a hybrid of European and Papuan lineages (prior to the Denisovan Papuan mixing). This can be further evaluated by comparing the deviance of the rooted triple (F,(H,P)) from our best model's expectations. The observed value, expected value and signed $X^2$ deviance are, respectively, HP: 15601, 15842.7, -3.7, FP: 13604, 14142.0, -20.5 and FH: 15580, 16153.3, -20.3. Thus in this best model there is no specific evidence for the pattern FH being overexpressed, indeed all these partial site patterns are under expressed and the degree of under expression of FP versus FH is no different. In this, the mixing of moderns with Neanderthal and Denisova, by itself, seems to explain why, in the raw data, the pattern FH appears to be overexpressed relative to FP.

It is also interesting to note that of the most deviant patterns in Table 7, quite a few can potentially be explained by sequencing / assembly errors. The most likely convergent patterns due to evolution would be changes in two long edges coming down from the ancestor. In this data all terminal edges are long, with the chimp being by far the longest. This makes patterns such as all but one of the ingroup sequences about equally likely by substitution. The most likely independent data errors would probably be between pairs of the most poorly sequenced genomes.

Interestingly, if we fix all the g values (tree edge lengths), but reoptimize the proportions of the four trees, allowing them to be independent (hence, one more parameter over the P(N) + P(D) model, but not nested within it) the deviance improves considerably, down ~100 units to 1714.0. The new proportions for the four trees in figure 9 are 0.9069, 0.0001, 0.0485 and 0.0446, respectively, versus the approximate proportions of 0.9022, 0.0422, 0.0531 and 0.0025 expected under the alternative previous model. This new model might suggest that there was not a common mixing of all out of Africa people with Neanderthals, but rather that the Han and French got their Neanderthal-like genes via a mixing with Neanderthals, perhaps in middle Eurasia, but the Papuans got their higher proportion of archaic genes solely via a mixing with a Denisovan like-creature (although how far its lineage might have been from the common ancestor of Denisova and Neanderthal is unclear). The model itself is also not valid, since if the gene mixing were independent events, then we still need four trees, with tree 1 being modified into having Han + French closest and Tree 2 having a cluster of Neanderthal + Han + French. The suggestion of this relaxed but not strictly valid model is directly testable, and assuming that there were two waves of moderns into Eurasia, with the first wave being the ancestors of Andamanese and Papuans, but the archaic Neanderthal mixing appearing later, then there should be a deficit of archaic genes in Andamanese (in particular Neanderthal genes).

### 3.11 Switching to finite sites models via a Hadamard conjugation

The finite sites models used here follow the methods of Waddell (1995). They are based on the infinite sites spectrum, transformed with an invariant sites plus gamma distribution Hadamard conjugation (Steel et al. 1993, Waddell 1995, Waddell et al. 1997). The parameters to be optimized are the parameters of the finite sites spectrum used above, allowing a single mixing of Neanderthals with the ancestors of the out of Africa people. The informative site spectrum produced by the infinite sites hierarchically structured coalescent model with mixing is then imported into the corresponding entries of the Hadamard gamma vector. The gamma vector is a model of all the bipartition weights in infinite sites space. The entries for the gamma vectors corresponding to the split of one taxon from all the others, e.g., $\gamma C$, are entered as free parameters. In order to balance these against the informative infinite sites fraction of the spectrum, which is not on a mutation scale yet, there is a free to optimize scale parameter. Finally, there is a free parameter $a$, for the fraction of invariant sites in the sequence and a shape parameter $k$, for the shape of the gamma distribution of site mutation/substitution/error rates. Thus, we have moved from a model with 8 free parameters to one with an additional $8 + 1 + 2 = 11$ free parameters, to give a total of 19 parameters to be optimized. Again, minimizing the total $X^2$ deviance is the objective of the numerical optimizer.

Before we look at the results it is useful to look at the predicted quality of the error rates in the various genomes. Estimates of sequencing error rate are made in Reich et al. (2010) table S2.4. These are, for transversions only, Denisova 0.000127, Neanderthal 0.000940, San 0.001523, Yoruba 0.001518, French 0.001008, Han 0.0015780, and Papuan 0.003159.

The solutions to this finite sites single Neanderthal mixing model are shown in table 8. There are four variants of the model shown, the first fits all site patterns, including uninformative sites, which include singletons. The second fits all site patterns, except the DP and DNP patterns, which misfit due to our not having the a gene infusion leading from the Denisovan lineage to the ancestor of the Papuan (due itself to the unanticipated retirement of Jorge Ramos at the peak of his population genetics career). The third focuses on fitting just the informative sites. This is a better approximation to a model where there are two site spectra being combined. This first spectrum is a finite sites spectrum due to a coalescent process and substitution, while the second is a independent error spectrum based on a star tree, with a unique edge length for each sequence, representing just the error rate in that sequence. Finally, there is this model minus the DP and DNP patterns.

The first model, labeled "Finite" in table 8, offers a better fit than the infinite sites model at 2393 verses 2554 $X^2$ units. This is due to the extra 11 free parameters and despite having to fit another nine site patterns. The parameter estimates are broadly similar to those seen earlier. The $\gamma$ values closely fit the relative proportions of the singleton site patterns in the data, since these have huge weight compared to the informative site patterns. Turing off the DP and DNP signals, the model labeled "Finite –DP" in table 8 improves considerably, and is again better than without the Hadamard conjugation, with a fit of 2045 verses a fit of 2179 without. The two most notable parameter changes are g6 becomes longer. This is because the single Neanderthal mixing is no longer being pressed by the high frequency of the DP and DNP patterns to explain them with a lineage branching closer to the common ancestor of D and N.

Table 8. The optimal fit and parameter values for the finite site models fitted with one mixing of Neanderthals with the out of Africa lineage.

| Model | Finite | Finite –DP | Finite –Unif | Finite –Uinf –DP |
|---|---|---|---|---|
| $X^2$ Fit | 2393.0 | 2044.7 | 1165.5 | 932.8 |
| g1 | 0.025 | 0.029 | 0.036 | 0.036 |
| g2 | 0.170 | 0.170 | 0.235 | 0.227 |
| g3 | 0.122 | 0.119 | 0.134 | 0.128 |
| g4 | 0.382 | 0.384 | 0.456 | 0.439 |
| g5 | 0.285 | 0.281 | 0.349 | 0.329 |
| g6 | 0.134 | 0.259 | 0.083 | 0.236 |
| g8 | 0.000 | 0.000 | 0.062 | 0.000 |
| P(N) | 0.067 | 0.055 | 0.021 | 0.019 |
| $\gamma$ C | 0.00270 | 0.00270 | 0.0054 | 0.0048 |
| $\gamma$ D | 0.00017 | 0.00017 | 0.0039 | 0.0023 |
| $\gamma$ N | 0.00028 | 0.00028 | 0.0068 | 0.0062 |
| $\gamma$ S | 0.00024 | 0.00024 | 0.0066 | 0.0060 |
| $\gamma$ Y | 0.00027 | 0.00027 | 0.0075 | 0.0068 |
| $\gamma$ F | 0.00026 | 0.00026 | 0.0079 | 0.0074 |
| $\gamma$ H | 0.00029 | 0.00029 | 0.0077 | 0.0072 |
| $\gamma$ P | 0.00030 | 0.00030 | 0.0089 | 0.0065 |
| k | 178.480 | 385.982 | 0.474 | 0.554 |
| a | 0.000 | 0.000 | 0.050 | 0.008 |
| Scale Factor | 0.995 | 0.989 | 56.654 | 47.562 |

Major gains in the overall fit appear when the uninformative sites in the data are ignored, as in the model labeled "Finite –U" in table 8, so that singleton gamma values are effectively acting more like free per sequence error rates if we had explicitly added in a star tree spectrum to model these values. In this case, they no longer closely match the terminal edge lengths, or the singleton site pattern frequencies, having become 10 or more times larger in some cases. They now more closely resemble the sequencing error rates in the sequences estimated by Reich et al. (2010). That is very low in Denisova, higher in Neanderthal and San, but highest in the genomes of Yoruba, French, Han and Papuan. This model now fits much much better, with an improvement of fit of about 1000 $X^2$ units! The contrast of this with the two previous "substitution only" models is also huge, with an improvement of about ~1000 units versus ~200 units, suggesting most of the problem is not substitutions, but data errors. The population genetic parameters have shifted noticeably. The inbreeding lengths of all the edges in the coalescent-based primary species trees have increased markedly, to values that are on average higher than any seen in the previous analyses. The inbreeding edge g6 has become shorter again as it is helping to explain the DP and DNP patterns now back in the model. The fraction of Neanderthal mixing has more than halved to around 2%. Additionally, the invariant sites and gamma shape parameters have taken values that suggest the rate of convergent events due to the finite sites model (which is now modeling what is apparently mostly sequencing /assembly error) is highly variable, with much higher rates at some sites. Finally, turning off the DP and DNP patterns again, there is the anticipated improvement in fit of around 200 units, down to below 1000 $X^2$ units. The inbreeding edge length g6 has increased again to around 0.24. Also, the error rate implied by the free $\gamma$ values has also reduced as, particularly on the D and P sequences, as they are no longer being called upon to help explain the high frequency of the DP and DNP patterns.

The major residuals of these four finite sites models are shown in table 9. A very interesting feature of the third and fourth models is that site patterns that are most likely to be due to sequencing errors have, generally, either dropped off the list or have become deficient in their observed frequency. In the very best fitting model, the pattern NSYFHP has now appeared as that which is most overexpressed in the real data. This pattern is consistent with the hypothesis that Denisovans may have picked up genes from an even more archaic human such as *Homo erectus*. It is also is consistent with the "Ron Jeremy's Grandfather" hypothesis, that this event may have been dominated by gene flow from the more archaic males into the more modern females, as suggested by the CA analysis earlier. Other patterns that are still overexpressed include the FHP pattern, which suggests that the inbreeding measure g2 wants to become larger but is being held back by other features of the data. These may be the next two most overexpressed patterns, namely YFH and SYFH, which are antagonistic to the parameter g2. The suggestion here seems to be that these individuals share something derived that the Papuan sequence does not, or looking at it the other way around, the Papuan may have some very archaic alleles, perhaps a gene injection from the archaic hominids thought to have been living in Indonesia until relatively recently. These include *Homo floresiensis*, which is may be a dwarfed descendant of *Homo erectus* that apparently lived until less than 20,000 years ago in the very region the Papuan's ancestors had to migrate through approximately 60,000 years ago.

Table 9. The thirty most deviant site patterns for the four finite site models fitted with one mixing of Neanderthals with the out of Africa lineage. They are, left to right, fitted to the full site spectrum, the full site spectrum minus the DP and DNP patterns, the full site spectrum minus the uninformative site patterns and, finally, the former minus the DP and DNP patterns also. For each the derived site pattern, the expected frequency predicted by the model and the signed $X^2$ statistic is shown.

| Model | F | | | F-D | | | F-U | | | F-U-D | | |
|---|---|---|---|---|---|---|---|---|---|---|---|---|
| Pattern | Expt. | $X^2$ | | Expt. | $X^2$ | | Expt. | $X^2$ | | Expt. | $X^2$ | |
| FHP | 4347.1 | 226.8 | NH | 789.4 | 227.3 | NSYFHP | 4069.4 | 122.7 | NSYFHP | 4142.2 | 97.0 | |
| NH | 809.4 | 201.2 | FHP | 4387.1 | 207.0 | DNP | 866.3 | 72.0 | FHP | 4744.1 | 74.8 | |
| DP | 726.4 | 163.5 | NP | 790.1 | 184.6 | FHP | 4765.2 | 69.3 | SYFH | 1330.1 | 73.6 | |
| NP | 810.1 | 161.7 | NF | 819.9 | 145.3 | DP | 841.2 | 62.8 | YFH | 1776.5 | 46.5 | |
| DNP | 787.1 | 137.5 | DSYFHP | 4832.1 | -120.5 | SYFHP | 8580.8 | -55.5 | SYFHP | 8491.6 | -42.5 | |
| NF | 834.5 | 130.9 | SF | 2025.8 | 81.5 | SYFH | 1371.2 | 53.9 | FP | 5038.2 | -37.9 | |
| DSYFHP | 4816.3 | -116.0 | NFHP | 812.1 | -78.9 | DF | 812.4 | -48.9 | NY | 1327.1 | -29.0 | |
| DNFHP | 1323.4 | -84.0 | DNFHP | 1299.1 | -73.6 | FP | 5080.3 | -45.2 | SFH | 1252.6 | 26.9 | |
| SF | 2022.9 | 82.7 | SYFH | 1346.2 | 65.4 | YFH | 1783.5 | 44.1 | NSYFH | 402.3 | 24.7 | |
| SYFH | 1354.7 | 61.4 | YF | 2790.1 | 44.1 | YP | 2871.5 | -42.3 | NS | 1612.2 | -23.2 | |
| NFHP | 746.1 | -46.9 | NY | 952.1 | 33.6 | DH | 787.5 | -28.0 | YP | 2771.6 | -22.3 | |
| YF | 2792.8 | 43.4 | NFP | 472.8 | -33.5 | SFH | 1252.5 | 26.9 | NP | 1025.9 | 20.8 | |
| NFP | 473.0 | -33.6 | YFH | 1828.4 | 30.3 | DNYFHP | 2218.9 | 25.3 | NSFH | 255.7 | 18.8 | |
| SYP | 1214.7 | -33.2 | SYP | 1196.4 | -27.8 | NH | 1054.8 | 23.7 | DNYFHP | 2251.4 | 18.6 | |
| NY | 959.3 | 30.7 | DYFHP | 1031.1 | -26.4 | SF | 2206.7 | 23.0 | HP | 5728.7 | -17.1 | |
| DYFHP | 1037.0 | -28.2 | YH | 2656.8 | 24.5 | SP | 2188.7 | -22.7 | SYP | 1151.7 | -16.5 | |
| YFH | 1839.5 | 27.4 | SFH | 1265.4 | 23.0 | NY | 1300.9 | -22.2 | NFHP | 661.5 | -15.9 | |
| DN | 12406.1 | -25.0 | DNF | 800.0 | 22.8 | HP | 5754.9 | -20.0 | SYFP | 1345.8 | -14.9 | |
| DFHP | 639.4 | -25.0 | DN | 12349.8 | -20.3 | SYP | 1163.2 | -19.1 | NH | 1090.1 | 13.9 | |
| DF | 744.0 | -23.1 | SYFHP | 8300.2 | -20.2 | NSFH | 255.4 | 18.9 | DSYFHP | 4309.6 | -13.4 | |
| SFH | 1267.6 | 22.4 | SH | 1940.4 | 19.7 | YF | 2914.6 | 17.6 | DY | 983.9 | 13.0 | |
| DFH | 417.6 | -22.4 | NYFHP | 1076.4 | -19.6 | NS | 1582.7 | -16.9 | SF | 2261.4 | 12.9 | |
| DNYFP | 519.9 | -22.0 | DFHP | 619.1 | -18.2 | SYFP | 1349.8 | -15.7 | DF | 706.2 | -12.3 | |
| YH | 2677.8 | 20.5 | DNYFP | 508.9 | -18.1 | NSYFH | 421.2 | 15.5 | DYFHP | 974.4 | -12.1 | |
| DNSFH | 501.5 | -19.8 | NHP | 507.4 | -17.6 | DNSFHP | 1704.0 | 15.4 | SHP | 1354.3 | 11.3 | |
| DNFH | 570.0 | -19.3 | NSFP | 271.9 | -17.0 | DNSFH | 478.9 | -12.3 | DNSFHP | 1731.4 | 10.5 | |
| NSFP | 275.7 | -18.6 | DSHP | 287.7 | -16.9 | SHP | 1352.2 | 11.7 | DNSFH | 471.1 | -10.1 | |
| SFHP | 2363.1 | 18.3 | DNSFH | 491.4 | -16.3 | DYFHP | 967.9 | -10.7 | DNYFP | 482.3 | -10.0 | |
| DSHP | 289.8 | -17.8 | SYFP | 1348.8 | -15.5 | DNYFP | 481.7 | -9.8 | NSFP | 253.6 | -9.7 | |
| SH | 1950.5 | 17.6 | FH | 4855.8 | 15.1 | DNS | 1797.3 | -9.6 | DNS | 1793.6 | -9.1 | |

## 3.12 Fitting Chromosome X

The most distinctive feature of chromosome X is its sex-linked character. Most males carry one copy of this chromosome, while most females carry two copies (although in extinct species such as *Homo sovietus*, this was never always clear). Thus if we fit chromosome X site patterns to the parameters estimated for the autosomes, then under a homogeneous model, the main difference will be the effective population size of the genes. If we call this ratio z, then $z = \frac{2N_m + 2N_f}{N_m + 2N_f}$, letting x be the effective proportion of males (that is $\frac{N_m}{N_m + N_f}$), then $x = 2 - \frac{2}{z}$ (and z must be in the range of 1 to 2). Fitting to the corrected minimum $X^{2"}$ model parameters with z = 4/3 so that x is 0.5 (or $N_m = N_f$), the fit of the 2253 site patterns from chromosome X has total deviance 288.6. When z is chosen to minimize $X^{2"}$ (at deviance value 275.7), then it takes on value 1.2013. This corresponds to a value of $N_m$ of nearly exactly 1/3 of the total effective population size of males and females. An approximate 95% confidence interval on z (assuming an inflation factor of $X^{2"}$ 1.5 due to inter-chromosomal variability) is obtained when $X^{2"}$ achieves a value of 6 above its minimum, which yields z in the range 1.2906 to 1.1149, which translates to x in the range 0.2061 to 0.4503. Switching to the robust fit criteria of minimum percent deviation, the value of z is 1.2317. Thus there seems to be good evidence to suggest that the long term effective population size of males is considerably smaller than that of females due to the "Genghis Khan" effect (that is, some males leaving many more offspring than others, and, therefore causing a larger fraction of males to leave no genes in future generations).

In terms of fit, the worst fitting patterns on chromosome X from the above fitting are SYF (signed deviance +32.7), DNSF +13.1, YFP +11.1, NSYFHP -10.2, DNSP +8.4, FHP +8.2, DNFH +7.1, YFHP +6.8, -DSYHP 6.2, and -FP 6.0. In contrast to the autosomes, none of the prime archaic interbreeding patterns show up as substantially over expressed on X. Many of the deviant patterns are unexpected, except for pattern NSYFHP which is under expressed and is compatible with the hypothesis that Denisova swapped genes with *H. erectus* but at a markedly lower rate than on autosomes, that is the "Ron Jeremy hypothesis."

Fitting chromosome X to the parameters of the minimum $X^2$ model with Neanderthal mixing we find the optimal fit of 281.62 with z = 1.134673 (1.049165-1.223907). This translates to x = 0.2374 but includes the previous value of z and x in the confidence interval. However, the hypothesis that $N_m = N_f$ is strongly rejected with p << 0.0001 even allowing for an inflation factor of 1.5. Switching off the deviance from the patterns DP and DNP makes little difference since they fit with the observed values on X with very well (again consistent with the "Ron Jeremy" hypothesis).

Finally, fitting with the same model as above but allowing for the finite sites model and a one cycle MS iterative correction, and modeling only the informative patterns, the optimal fit is 269.46 with z = 1.137797. This translates to x = 0.2422 and again includes the previous value of z and x in the confidence interval. The hypothesis that $N_m = N_f$ is still strongly rejected with p << 0.0001, again allowing for an inflation factor of 1.5.

**3.13 A note on P and D tests with finite sites models**

When dealing with the coalescent plus the infinite sites model, a rooted 3 species tree has some special properties similar to those ascribed to Buneman (1971) for the 4 species unrooted distance tree. These properties are that the dominant source of gene flow is indicated by which of the patterns 011, 101 and 110 is maximal. Under any hierarchical coalescent, without reticulate events (including within members of the same "species"), the two less frequent patterns should have equal expectation. That one is clearly maximal and indicated by the "species tree" may be tested using the $P_1$ test of Waddell et al. (2000, 2001). Whether the other two are equal may be tested with a binomial test, the P$_2$ test of Waddell et al. (2000, 2001). If this test is rejected then, assuming there is no data processing bias, it is an indication of introgression. Another version of this test, useful with frequency counts greater than 5 or so, is to use an $X^2$ statistic. Using the techniques above, this can be recalibrated into a scaled $X^2$ statistic to take into account the effect of genetic linkage plus other unbiased fluctuations as might be caused by selection, etc. If f[011] > f[101] and f[110] then, assuming no other violations of the basic model, the excess of f[110] over f[101] indicates the degree of introgression between lineages 2 and 3. As a proportion this is f[110]-f[101]/(f[110]]+f[101]). This latter quantity used by Reich et al. (2010) for example, is thus intimately related to the tests of Waddell et al. (2000, 2001). Obviously, what works for a triplet can be extended to predict a larger structure, the basic principal behind Split Decomposition, for example (Bandelt and Dress 1992). It can also be readily extended to reticulate diagrams for populations, when it is recognized that these decompose into a set of weighted trees, with membership equal to $2^x$, where $x$ is the number of reticulate events (Waddell 1995).

To extend the P and D statistics to finite sites models is highly desirable. A robust approach to do this was developed in Waddell (1995) and is redescribed here. The critical step is to use a Hadamard conjugation (a form of discrete fast Fourier transform, Hendy and Penny 1993) to make corrections between sequence space and the idealized infinite sites or tree space. When dealing with lineages that coalesce relatively quickly compared to the number of new mutations per lineage, then the Hadamard conjugation is a convenient approximation at intermediate probabilities of parallelism or convergence. When dealing with the complex mixtures of weighted trees generated by coalescent patterns, it is even more appealing as no more computational effort is needed. One apparent alternative is to discretize weighted tree space into segments, conjugate (transform) them separately and reassemble them, this will improve the approximation in some cases.

The conjugation on four species can be run either from sequence space to tree space, in which case the P-statistic can be likened to $\gamma 5 - \gamma 6$ in the above example. The significance of this difference, under an i.i.d. model, can be estimated by deriving the variance covariance matrix (Waddell et al. 1994) of the transformed data. This may be called a finite sites P$_2$ statistic. In the same way a finite sites D statistic may be defined as $\gamma 5 - \gamma 6 / (\gamma 5 + \gamma 6)$. This is fine in many circumstances as long as neither is substantially negative.

When other parameters of the model are desired to be optimized at the same time, it is often more convenient to run the model in the direction gamma to sequence space and use a deviance statistic to measure discord. Using the KL deviance statistic, then the method is ML under certain assumptions. With a non-instant coalescence at splits, a true ML solution requires discretization of tree space. One such approximation is described in Waddell et al. (2001). A close approximation to the ML solution is obtained via the Hadamard conjugation as described above. The significance of the result may be estimated using a likelihood ratio statistic (taking into account the fact that both $\gamma 5$ and $\gamma 6$ may have a boundary at zero enforced, Ota et al. 2000).

**3.14 Solving for ancestral population sizes**

It is possible to estimate approximate ancestral effective population sizes using the methods in Waddell (1995). These require an estimate of the duration of each internal edge of the tree in terms of the number of generations. Following this we have $x + 1$ unknowns and $x$ different equations. The extra unknown is the effective population size at the root of the tree. In Waddell (1995) this was dealt with with the assumption that the common ancestor of human, chimp and gorilla had the same effective population size as that of human and chimp. Here there are multiple descendant lineages, so where to put such a constraint is less obvious. The alternative is to plot out the estimated effective population size along internal edges of the tree for a range of values of the ancestral population size (Nr).

Estimating the mean divergence times of populations given multiple sources of error, including divergence time calibration uncertainties, edge length errors, unknown ancestral population sizes, fluctuating mutation rates, deviations from a single tree spectrum due to differing gene trees, and model misspecification all make divergence time estimation particularly challenging (Waddell 1995). Firstly, the estimates of the duration of internal edges of the genome on the species gene tree in real time need to be made, then transformed into numbers of generations. With the current data, to make estimates in real time is complicated by the fact that there has been reticulation in the tree, most notably the Neanderthal and Denisovan introgression events. A cautious approach to this problem involves selecting specific pairwise distances over others that are expected to be more distorted by reticulate events. There is also an issue of appreciable quantities of sequencing error, although the table S6.2 of Reich et al. (2010) makes a correction for these to pairwise distances. The divergence of Neanderthal and Denisova genes is set at 644k years ago (apparently unaffected by gene flow with later species, but perhaps distorted by earlier gene flow). Africans split from these archaics about 810k ago (based on the average of the YN, YD, SN and SD distance pairs). San split from other humans about 593k ago (using the SY distance). The split of Yoruba from non-Africans is set at about 520k ago (a bit less than the YF and YH pairs to compensate for the Neanderthal gene inflow). The split of Papuan and Han is set at 403k, but this is almost certainly biased upwards by up to 10% of archaic genes, and is barely older than the split of the French and Han pair. A more realistic number for HP might be 380k. It seems safer to set the earlier F(HP) time using the (very) approximate 20,000 year gap in evidence of modern human in western versus eastern Asia based on archaeology.

Next, it is necessary to estimate the average generation time. Waddell (1995) gives a range of generation times for modern humans of 18 to 25 years. Recent assessments of genealogies by Matsumura and Forster (2008) indicate generation times for Eskimos closer to 30 years and chimps at 19 and 24 years. There is evidence that earlier hominin populations had relatively few members surviving over 40 years (Trinkaus et al. 2011, who identify 45 versus 14 archaic adults under and over 40, while upper Paleolithic moderns are reported at 36 versus 13, for a Fisher's exact 2 tailed test result of p = 0.82 or even more homogeneous than the original authors noted). Thus it would seem appropriate to take a mean generation time estimate of earlier humans in the range of 20 to 30 years, with 25 years as perhaps a reasonable mean. Thus we arrive at durations of the internal edges of the species tree at $t_5$ = 166,000 years or 3320 generations, $t_4$ = 220,000 / 4400, $t_3$ = 60,000 / 1200, $t_2$ = 130,000 / 2600 and $t_1$ = 25,000 / 500.

The hierarchical coalescent models edge lengths are measured on units of generations divided by twice the effective population size (for autosomes in a diploid). We need to calculate how these relate to gene splitting times. Thus we come to a series of equations. For example, the difference in the splitting times of genes across the root node versus the ancestral node of modern humans is equal to $N_r + g_4N_4 - (p_4c_4N_4 + (1-p_4)N_r + (1-p_4)g_4N_4)$ or $-p_4c_4N_4 + p_4(N_r+g_4N_4)$. Here $p_4$ is the probability of coalescing on edge 4, that is $p_4 = (1 - e^{-g_4})$, while $c_4$ is the expected time it takes to coalesce along edge 4 which is equal to $c_4 = (p_4 - g_4e^{-g_4})/p_4 = ((1-e^{-g_4}) - g_4e^{-g_4})/(1-e^{-g_4})$. At the other extreme, for genetic divergences towards the tips of out species tree, the difference of distances measured at the node above and below edge 1 is equal to $p_1c_1N_1 + (1-p_1)p_2(c_2N_2+g_1N_1) + \ldots$ . While the distance measured across the node below are equal to $p_2(c_2N_2+g_1N_1) + \ldots$ . These alternating terms mostly cancel when the first term is subtracted from the second term to give $d_2 - d_1 = -p_1c_1N_1 + p_1p_2(c_2N_2+g_1N_1) + p_1(1-p_2)p_3(c_3N_3+g_2N_2+g_1N_1) + p_1(1-p_2)(1-p_3)p_4(c_4g_4+g_3N_3+g_2N_2+g_1N_1) + p_1(1-p_2)(1-p_3)(1-p_4)(N_r+g_4N_4+g_3N_3+g_2N_2+g_1N_1)$. This series of non-linear equations is best solved simultaneously as, logically, each solution of an ancestral population size at a higher node affects the values for each lower node.

The numbers used above yield edge length durations on the main species tree of approximately ((D,N):164:0.267,(S,(Y,(F,(H,P):20:0.027):120:0.171):73:0.093):217:0.342)

(where the first number is the duration of the edge in thousands of years and the second number is the g edge length). Assuming an ancestral population size of 20,000 gene copies, then solving these equations, the inference of the ancestral gene population sizes on the tree turn out to be ((D,N):15,400,(S,(Y,(F,(H,P):18,600):17,500):19,600):15,800), under these varied assumptions (that we have moderately accurate relative divergence times, the assumed generation times are correct and the assumptions used in estimating $g_1$-$g_5$). Even taking the most distinct values of $g_1$-$g_5$ obtained under the best fitting $G^{2''}$ model of 0.02956, 0.17743, 0.09573, 0.35100 and 0.27947, the estimated population sizes are little different at ((D,N):14,700,(S,(Y,(F,(H,P):16,900):16,900):19,100):15,500).

These ancestral population size numbers are all surprisingly similar, which is hard to reconcile with the out of Africa event, for example (which here is shown as a minimal reduction in effective population size from 19,600 to 17,500). The complex mixing of genes near the out of Africa event, along with some relatively short and poorly estimated edge lengths (such as that of HP) that might be off by a factor of 2 or more (if modern humans expanded from west to east Asia in 10,000 years of less as is possible given the vagaries of the archeological evidence) and also potentially high data error rates. The more ancient parts of the tree would seem more likely to be stable to these effects, so that the populations of modern humans and the archaic out of Africa (N and D) would seem to be roughly similar.

## 4 Discussion

Overall, the fitting shows that a hierarchical structured coalescent model with at least two introgression events between archaic humans and out of Africa Moderns leads to a substantial increase in fit. Overall fit however, is still far far worse than could be expected.

It seems that to improve the fit a number of factors may come into play. Firstly, there are too many private NH, NF and NP patterns. Secondly, the latter of these, NP, seems markedly less than the former two. Thirdly, there may be too many sequencing/alignment errors in the present data to confidently move towards refining so many parameters and the overall fit. The marked improvement in fit when a finite sites model is employed is consistent with this.

One model that may do a better job of describing the data with fewer parameters is independent mixing of Neanderthal genes with Han and French, but to a nearly identical total degree. Also, lesser mixing of Neanderthal genes into Papuan, made up for by a larger proportion of archaic alleles in Papuans coming from the mixing with an archaic that is only slightly closer to Denisova than to Neanderthal. This would in turn suggest that the mixing with Neanderthals was not purely right out of Africa and it was not a single event. Instead, there may have been opportunity for European ancestors to pick up Neanderthal alleles, in the unknown part of Eurasia they existed in prior to moving into Europe, ditto and independently for the ancestors of the East Asians, while Papuan ancestors moved fairly rapidly through the zone of classical Neanderthals and picked up most of their archaic genes in the Indonesian region. The form of this ancestral population may have been about equally related to Neanderthals and Denisovans, but may also have had an appreciable proportion of even earlier (e.g., *Homo erectus* genes) in its genome. This last point comes up in a number of analyses including the resampled NeighborNet and the finite sites model, but confirmation is difficult as the rate of sequencing / assembly error could be having a similar effect.

Closing with reference to the quotations of Darwin in the introduction, it is interesting to ponder further the question of which human lineages are different species. In almost all ways, modern human lineages meet the definitions of being at most different races of the same species, and some would argue much closer to simply different populations. In terms of time since origin, genetic distances, and fertile interbreeding, modern humans clearly constitute what any biologist would call a single species. The question of whether Neanderthals and Denisovans should be considered the same species is less clear, but again many would argue that they too might be included in the same species. However, by one species criteria little considered, we may not always behave as one species. This is the rate and / or amount of gene flow between hominid populations when they become sympatric. It seems that in the case of archaic hominids the out of African moderns were initially a very small group (perhaps with an effective population size of less than 1000) that managed to replace all the archaic hominids in Eurasia (probably numbering at least in the tens of thousands, and supposedly fairly well adapted to their habitats). That only 2-5% of archaic genes survived genes overall, and this potentially restricted to a relatively few matings early on, suggests a real whitewash of the archaics. It is very unclear if this is the behavior excepted of the same species genes when different populations come into contact.

Fortunately, large scale genomic sequencing and advancing methods of phylogenetic-population genetic analysis should allow a clear answer to the question of how typically populations of *Homo* act in terms of gene flow in comparison to other Primate species in particular. Ominously, the replacement of archaic hominids with minimal interbreeding does not seem so far from what has happened in our own recent history, for example when cultured white Australians encountered a relatively large stable population of Tasmanian Aborigines and completely replaced them by means of war and disease in a few generations with almost no interbreeding. It is interesting to ponder if that is the behavior expected of members of the same species, or if we need to recognize more clearly the genetic ways in which we do not behave as members of the same species. Darwin himself seemed very unsure on this question, and we at last have the opportunity to look at it in unflinching detail.

## Acknowledgements

This work was supported by NIH grant 5R01LM008626 to PJW. Thanks to Dick Hudson, Nick Patterson, David Bryant, Martin Kircher, Hiro Kishino and Dave Swofford for helpful discussions and assistance with software.

## Author contributions

PJW originated the research, developed methods, gathered data, ran analyses, interpreted analyses, prepared figures and wrote the manuscript. XT implemented methods in C and PERL, ran analyses, prepared figures, interpreted analyses and commented on the manuscript. JR implemented the 7-taxon single tree hierarchical coalescent calculations for the infinite sites model in Excel.